\documentclass[aps,showpacs,showkeys]{revtex4}
\usepackage[usenames]{color}
\usepackage{epsf}
\usepackage{epsfig}
\usepackage{amsmath}
\usepackage{graphicx}
\begin{document}
\baselineskip 16pt
\title{{\bf Singularity structure of the $\pi N$ scattering amplitude in a meson-exchange model
up to energies $W \leq 2.0$~GeV}}
\author{L. Tiator$^a$, S.S. Kamalov$^{a,b,d}$, S. Ceci$^c$, G.Y. Chen$^d$, D.
Drechsel$^a$, A. Svarc$^c$, and S.N. Yang$^d$\\
$^a$Institut f\"ur Kernphyik, Universit\"at Mainz, D-55099 Mainz,
Germany\\$^b$Bogoliubov Laboratory for Theoretical Physics, JINR,
Dubna, 141980 Moscow Region, Russia\\$^c$Rudjer Boskovic Institute,
Division of Experimental Physics, HR-10002 Zagreb,
Croatia\\$^d$Department of Physics and Center for Theoretical
Sciences, National Taiwan University, Taipei 10617, Taiwan}
\date{\today}
\begin{abstract}
Within the previously developed Dubna-Mainz-Taipei meson-exchange model, the
singularity structure of the $\pi N$ scattering amplitudes has been
investigated. For all partial waves up to $F$ waves and c.m. energies up to $W
\sim 2$~GeV, the $T$-matrix poles have been calculated by three different
techniques: analytic continuation into the complex energy plane, speed-plot and
the regularization method. For all 4-star resonances, we find a perfect
agreement between the analytic continuation and the regularization method. We
also find resonance poles for resonances that are not so well established, but
in these cases the pole positions and residues obtained by analytic
continuation can substantially differ from the results predicted by the
speed-plot and regularization methods.
\\
\end{abstract}
\pacs{11.80.Gw, 13.75.Gx, 14.20.Gk, 25.80.Dj}
\keywords{pion-nucleon interaction, baryon resonances, T-matrix poles,
speed plot, regularization method}

\maketitle
\section{Introduction}
\label{sec1}
Ever since the $\Delta(1232)$ resonance was discovered by Fermi and
collaborators in 1952~\cite{Anderson:1952xx,Fermi:1952zz,Anderson:1952nw}, the
excitation spectrum of the nucleon has played a fundamental role in our
understanding of low-energy hadronic physics. The most direct evidence for
resonance structure is based on pion-nucleon elastic and charge-exchange
scattering. Because total angular momentum, parity, and isospin are conserved
within the realm of the strong interaction, the $S$ matrix for the reactions
$\pi + N \rightarrow \pi' + N'$ may be decomposed into the partial wave
amplitudes $T^I_{\ell \pm}$, with $I$ the isospin, $\ell$ the orbital angular
momentum, and the $\pm$ indicating the total spin of the hadronic system,
$J=\ell \pm {\textstyle {\frac {1}{2}}}$. Two decades after the discovery of
the $\Delta(1232)$, a dedicated program at the meson factories had provided
enough data to establish a rich resonance spectrum of the
nucleon~\cite{Donnachie:1972dh,Donnachie:1973xx}. The further partial wave
analysis was driven by studies of the Karlsruhe-Helsinki
(KH)~\cite{Hohler:1979xx,Koch:1980ay} and Carnegie-Mellon--Berkeley
(CMB)~\cite{Cutkosky:1979fy,Cutkosky:1979zv} collaborations. In the following
years, R. Arndt and coworkers at VPI compiled the data base
SAID~\cite{Arndt:1985vj}, which was later updated and extended in collaboration
with GWU~\cite{SAID04,Arndt:2006bf}. The works of the CMB, KH, and GW/VPI
groups are the main sources of the nucleon resonance listings in the Review of
Particle Physics (PDG)~\cite{PDG2008}.
\\

In the most intuitive way, a resonance is an intermediate state of target and
projectile that lives longer than in a typical scattering process. Translated
into the language of scattering theory, resonances are defined as poles of the
$S$ matrix on unphysical Riemann sheets. Different methods were developed to
derive the resonance properties from the observables. In the 1930's it was
suggested that a Breit-Wigner function should be a good representation for a
resonance pole, and the Breit-Wigner formula for spin zero particles and its
generalization to finite spin were developed (see an illustrative discussion in
Cottingham and Greenwood, p.241 in Ref.~\cite{Breit-Wigner}). Later the
discussion centered on the rapid increase of the eigenphase shifts through
$90^{\circ}$ and on the related backward looping of Argand
diagrams~\cite{Dalitz:1970xf}.
\\

However, Breit-Wigner parametrizations have been found to be very model
dependent. As was recently shown in the framework of effective quantum field
theory, Breit-Wigner masses are in general field-redefinition
dependent~\cite{Djukanovic:2007bw}. The same model dependency also applies to
electromagnetic properties as charges, magnetic moments, transition moments,
and form factors. On the other side, all these resonance properties are
uniquely defined at the pole of the $S$ matrix~\cite{Gegelia:2009py}.
\\

The analytic properties of the $S$ matrix are imposed by the principles of
unitarity and causality. Because of unitarity, each physical channel leads to a
square-root branch point of the partial wave amplitude $T(W)$ at the respective
threshold, with the result that $T(W)$ is a multi-valued function in the complex
$W$ plane. In particular, a partial wave for elastic $\pi N$ scattering is
described by the amplitudes  $T^{[1]}$ on the physical and $T^{[2]}$ on the
unphysical sheet. The experimental amplitudes are identified with the
amplitudes above the cut, $T_{\rm {exp}}(W)= {\rm {lim}}_{\epsilon \rightarrow
0}\,T^{[1]}(W+i\,\epsilon)$, with $W > M+m$ and $\epsilon > 0$. As a
consequence, the physical sheet has a discontinuity over the real axis, along the
right-hand cut $M+m \leq W < \infty$, with $M$ the nucleon and $m$ the pion
mass. In addition, a relativistic theory has a left-hand cut starting at $M-m$.
Causality requires that the physical sheet be free of any further singularity,
the nucleon resonances should appear as simple poles on the unphysical sheet
closest to the real axis of the physical sheet, in agreement with H\"ohler's
remark~\cite{Hoe2001}: \textit{``It is `noncontroversial among theorists' (see
Chew \cite{Chew1976} and the references in my `pole-emics', p.697 in Ref.
\cite{Hoe2000}) that in S-matrix theory the effects of resonances follow from
first order poles in the 2nd sheet.''} A pole on the second sheet, described by
$T^{[2]}(W)\approx r_p/(M_p-W-i\,\Gamma_p/2)$, will often lead to a maximum
of the experimental cross section near $W=M_p$. The  resonance is therefore defined
by (i) its pole position in the complex c.m. energy plane at $W_p=M_p - i\,
\Gamma_p/2$, with $M_p$ the real part of the pole position and $\Gamma_p$ the width of
the resonance, and (ii) the residue of the amplitude, $r_p = |r_p|
{\rm {exp}} (i\, \theta_p)$, at the pole.
\\

The focus of the present work is on how to extract resonance properties from
the data, that is, how do we find a pole in the complex energy plane having at
our disposal only data on the real axis and use different pole-extraction
methods to extract one, and only one pole per resonance. The problem is how to
eliminate the background from the experimental amplitude and to isolate the
pole contribution. The time delay method and the related speed plot technique
turned out to be the most quantitative tools to detect poles and extract their
parameters from the data. A resonance is characterized by a maximum time delay,
the time passing between the arrival of a wave packet and its departure from
the collision region. In general, a large time delay indicates the formation of
an unstable particle in the intermediate state. However, misleading effects can
occur by rapid variation of backgrounds, such as narrow cusp effects above
$S$-wave thresholds or spurious singularities due to phenomenological
parametrization of form factors and cut-offs. If a resonance lives long enough,
it should decay into all energetically possible final states, unless prevented
by general selection rules. Furthermore, the pole position derived from the
data should not depend on how the resonance is excited or decays. Whereas such
resonances exist in atomic and also in nuclear physics, some caveat is in order
for nucleon resonances. As an example, a simple classical model of the
$\Delta(1232)$ leads to the conclusion that the pion stays in its orbit around
the nucleon for only about $100^{\circ}$ of a full circle.
\\

The pole positions of nucleon resonances have been traditionally derived from
the data by the speed plot~\cite{Donnachie:1972dh,Hohler:1992ru}, which is
related to the time delay. The ``speed'' is defined by the slope of the
amplitude with energy, $dT(W)/dW$, which eliminates constant backgrounds. The
speed plot shows the modulus of the speed, $\mbox{SP}(W)=|dT(W)/dW|$, and
resonances are identified with peaks in the plot. The resonance parameters are
then obtained by fitting the speed of a single pole to the data at physical
values of the energy. The idea of the speed plot has been recently generalized
to higher derivatives by the ``regularization method''
(RM)~\cite{Ceci:2008zza}. Within a convergence circle given by the closest
neighboring singularity, the Laurent expansion about a pole is given by the sum
of $T^{{\rm {pole}}}(W)$ and a Taylor series $T^{{\rm {reg}}}(W)$. With an
increasing number of differentiations, the signature of the pole sticks out
more and more sharply, whereas more and more leading terms of the Taylor series
disappear. It goes without saying that the differentiation of the data will
fail after a few steps because of numerical instabilities. However, the
regularization method is an interesting tool to study the singularity structure
of analytic models. In particular, this method will reveal any rapid variation
due to cusp effects and spurious singularities introduced by phenomenological
parameterizations.
\\

The inherent model dependence of all the methods has been often overlooked.
Examples are the decomposition of the scattering amplitude in a Breit-Wigner
form and a background~\cite{Hoe2000} and the fact that eigenphases of a
multi-channel system usually do not run through $90^{\circ}$ because of the
Neumann-Wigner no-crossing theorem~\cite{Neumann-Wigner29}. With regard to the
latter point, Dalitz and Moorhouse~\cite{Dalitz:1970xf} stated: \emph{``With
such a complexity of branch cuts without physical significance, we must
conclude that  the eigenphase representation for the S matrix is not  generally
a useful representation for the scattering in the neighborhood of a
resonance''}. On the other hand, a single-channel procedure can not reveal much
information about resonances `far away' from that channel, that is, if the
particular resonance has only a small branching ratio for the particular channel.
\\

The dual nature of the nucleon resonances, the ``bare'' resonance based on
quark configurations and the ``dressed'' pion-nucleon continuum state, has been
studied by several collaborations within Lagrangian models, in particular by
EBAC at Jefferson Lab~\cite{Matsuyama:2006rp,JuliaDiaz:2007kz}, the groups at
Giessen~\cite{Shklyar:2004ba}, J\"{u}lich~\cite{Doring:2009yv},
Valencia~\cite{Inoue:2001ip}, and Darmstadt~\cite{Kolomeitsev:2003kt} as well as
the Dubna-Mainz-Taipei collaboration~\cite{Chen:2007cy}. Such approaches are
also the basis for structure investigations of resonances by photo- and
electroproduction.
\\

Although Lattice QCD (LQCD) has obtained promising results for the masses and
several resonances of the low-lying
hadrons~\cite{Alexandrou:2009xk,Walker-Loud:2009,Bula2010}, the large pion mass
used and the quenching approximation make it (yet) impossible to treat the
resonances as a pion-nucleon scattering state. As shown by
Ref.~\cite{Pascalutsa:2006up} for the $N \Delta$ form factors, the chiral
extrapolation from the stable $\Delta$ at large pion masses to the experimental
pion mass yields unexpected and rather dramatic non-analytic effects at the
$\Delta \rightarrow N \pi$ threshold. The very fact that lattice theory can not
yet describe the pion-nucleon final state interaction, makes it impossible to
compare the lattice data to the experimental scattering amplitudes in a direct
way. However, the LQCD data could be compared to the results of a dynamical
model obtained with the same pion mass as used in LQCD. As long as the pion
mass is large, $m_\pi > m_{N^*}-m_N$, the N* appears as an excited bound state.
In this case the mass predicted by LQCD should be compared to the "dressed"
mass of the dynamical model, which contains the "bare" mass and a real
self-energy due to meson loops. In the mass region where the N* decays, the
LQCD calculation would provide a complex amplitude, comparable to the
scattering phases of a partial wave expansion. Based on the work of
L\"uscher~\cite{Luescher:1986}, the width of the rho meson~\cite{Aoki:2007} and
of the Delta resonance~\cite{Bernard:2008ax} have been recently studied. In
such a case the lattice data can be treated like experimental amplitudes, that
is, by speed plot or Breit-Wigner analyses. However, as long as the LQCD pion
mass lies above the physical value, an extrapolation to the physical pion mass
will still be necessary by use of a dynamical model or an effective field
theory.
\\

In the present contribution we study the Dubna-Mainz-Taipei model
(DMT)~\cite{Chen:2007cy} by comparing the pole parameters resulting from
analytic continuation with approximate procedures such as the speed plot and
the regularization method. The DMT model is a field-theoretical meson-exchange
model for $\pi N$ scattering. In contrast to the partial wave amplitudes
obtained in
Refs.~\cite{Cutkosky:1979zv,Cutkosky:1979fy,Arndt:1985vj,SAID04,Arndt:2006bf},
which inevitably involve a smoothing of the data, the energy dependence of the
DMT amplitudes is largely determined from theoretical considerations even
though there are free parameters in the model. Another strong point of the DMT
model is the fact that the prescribed background contribution, a major source
of the model dependence for the resonance extraction procedure, has been found
to give excellent agreement if applied to low energy electromagnetic pion
production~\cite{Kamalov:2001qg}. In Sect. II we give an overview of the
speed-plot, time-delay, and regularization methods to derive the pole
parameters. The Dubna-Mainz-Taipei model is presented in Sect. III, in
particular with regard to the definitions of resonant vs. background terms as
well as form factors and cut-offs. Our results for the resonance parameters are
reported in Sect. IV, and the different techniques to derive these parameters
are compared. We conclude with a summary and outlook in Sect. V.\\
\section{How to find a resonance}
\label{sec2}
\subsection{Time delay}
\label{subsec2a}
In the framework of a potential scattering problem, resonance
phenomena are related to the formation and the decay of intermediate
quasi-stationary states. An ideal resonance should decay in all
energetically possible final states, unless forbidden by selection
rules for a specific channel. Provided that the interaction is of
short range, resonances can be characterized by the {\emph {time
delay}} between the arrival of a wave packet and its departure from
the collision region. In general, the time delay shows a pronounced
peak at the resonance energies. The time delay was introduced by
Eisenbud in his Ph.D.~thesis~\cite{Eisenbud:1948xx} and later
applied to multi-channel scattering theory by
Smith~\cite{Smith:1960zza}. Following the work of
Refs.~\cite{Wigner:1955zz,Dalitz:1970xf,Nussenzweig:1973hp,Suzuki:2008rp},
we define the time delay for a single-channel scattering problem in
a partial wave as follows:
\begin{equation}
\Delta t(W) = {\rm{Re}}\,\left(-i\frac{1}{S(W)}\frac{dS(W)}{dW}\right) = 2\, \frac {d \delta (W)}{dW}\,,
\label{eq:2.1.1}
\end{equation}
where $S(W)={\rm {exp}}[2i\delta(W)]$ is the $S$ matrix and $\delta(W)$ the scattering phase shift.
A simple ansatz for a unitary $S$ matrix with a pole at $W_p=M_p-\textstyle {\frac {i}{2}}\Gamma_p$
and a constant background phase $\delta_B$ is given by
\begin{equation}
S(W)=\frac{M_p+\textstyle {\frac {i}{2}}\Gamma_p-W}{M_p-{\frac {i}{2}} \Gamma_p-W}\,e^{2i\delta_B}= e^{2i \delta_R (W)}\, e^{2i\delta_B}\,,
\label{eq:2.1.2}
\end{equation}
where $\delta_R (W) ={\rm{\arctan}}[\textstyle {\frac {1}{2}}\Gamma_p/(M_p-W)]$
is the resonant phase. The related $T$-matrix $T=(S-1)/(2\,i)$ and the real
matrix $K=i\,(1-S)(1+S)^{-1}$ take the forms
\begin{eqnarray}
T(W)&=& \frac{{\frac {1}{2}}\Gamma_p}{M_p-{\frac {i}{2}}\Gamma_p-W}\, e^{2i\delta_B} + {\rm {sin}}\delta_B e^{i\delta_B}\,, \label{eq:2.1.2.a}\\
K(W)&=& \frac{{\frac {1}{2}}\Gamma_p+ (M_p-W){\rm {tan}}\delta_B}{M_p-W-{\frac {1}{2}}\Gamma_p {\rm {tan}}\delta_B\,} \,. \label{eq:2.1.2.b}
\end{eqnarray}
Combining Eqs.~(\ref{eq:2.1.1}) and (\ref{eq:2.1.2}), we obtain a simple Breit-Wigner form for the time delay,
\begin{equation}
\Delta t(W)=\frac{\Gamma_p}{(W-M_p)^2+ \textstyle {\frac {1}{4}}\Gamma_p^2} \,.
\label{eq:2.1.3}
\end{equation}
In this ideal case, the maximum time delay is $\Delta t(M_p)=4/\Gamma_p$.  We
observe that both the real ($M_p$) and imaginary ($-\textstyle {\frac
{1}{2}}\Gamma_p$) parts of the pole position are determined by the time delay
(and therefore also by the scattering matrix) at real (physical!) values of the c.m. energy $W$.\\

In order to describe a system of coupled channels, the lifetime matrix $Q$ was
introduced~\cite{Smith:1960zza},
\begin{equation}
Q_{ij}(W) = -i\frac{dS_{ik}(W)}{dW}\,S^{\ast}_{jk}(W)\,.
\label{eq:2.1.4}
\end{equation}
The analog of the time delay for a multi-channel system was found to be the
trace of $Q$ as function of $W$. This trace takes a more transparent form after
diagonalization of the $S$ matrix~\cite{Dalitz:1970xf,Haberzettl:2007ww},
\begin{equation}
{\mathrm {tr}} [Q(W)] = 2\,\sum_{\alpha} \frac {d \delta_{\alpha} (W)}{dW}\,,
\label{eq:2.1.5}
\end{equation}
with $\delta_{\alpha}$ the eigenphase shifts. However, a realistic
application of the lifetime matrix requires the  knowledge of all
open channels, that is, the reactions $\pi N \rightarrow \pi N$,
$\pi N \rightarrow \pi \pi N$, $\pi \pi N \rightarrow \pi \pi N$,
$\pi N \rightarrow \eta N$, and so on. As a consequence of Wigner's
no-crossing theorem~\cite{Wigner1929}, individual eigenphases have a
complicated energy dependence. It is therefore only the sum of the
eigenphases that shows distinct resonance structures. Based on this
observation, it has been recently proposed to search for resonance
parameters by studying the traces of multi-channel $T$ and $K$
matrices~\cite{Ceci:2006jj,Workman:2008iv}. The respective
scattering matrices were constructed from experimental data for the
$\pi N$ and $\eta N$ channels and models for the two-pion channels.
\subsection{Speed plot}
\label{subsec2b}
The {\emph {speed plot}} of a partial wave amplitude $T$ is defined by
\begin{equation}
\mathrm{SP} (W) = \left| \frac{dT(W)}{dW} \right|\,.
\label{eq:2.2.1}
\end{equation}
As was recognized by the Particle Data Group already in the early 1970's, the
speed plot is a convenient tool to extract the pole position of a
resonance~\cite{Rittenberg:1971xw,Donnachie:1973xx}. This technique was
intensively studied by H{\"o}hler who wanted to extract resonance parameters
from the Karlsruhe-Helsinki partial wave analysis
(KH80)~\cite{Hohler:1983xx,Hohler:1992ru}. Because the KH analysis was
restricted to elastic pion-nucleon scattering, a multi-channel treatment
like the construction of the lifetime matrix was out of reach. \\

For a single channel system, the time delay and the speed plot are
identical up to a factor 2. In particular, the ansatz of
Eq.~(\ref{eq:2.1.2.a}) leads to the speed
\begin{equation}
\mathrm{SP} (W) = \frac{{\textstyle {\frac {1}{2}}} \Gamma_p}
{(W-M_p)^2+{\textstyle {\frac {1}{4}}}\Gamma_p^2}\,.
\label{eq:2.2.2}
\end{equation}
As a result, the speed plot shows a maximum for $W=M_p$, which defines  the
real part of the pole position in the complex $W$ plane. The imaginary part of
the pole position can be obtained from the relation
\begin{equation}
\mathrm{SP}(M_p \pm {\textstyle {\frac {1}{2}}} \Gamma_p) ={\textstyle {\frac
{1}{2}}}\,\mathrm{SP}(M_p)\,. \label{eq:2.2.3}
\end{equation}
In practical applications, this straightforward method is numerically not very
stable and fails completely for the $S_{11}(1535)$ resonance, which is very
close to the $\eta$ threshold. In order to allow for inelasticities, we
therefore assume the functional form
\begin{equation}
T(W) = \frac{r_p}{M_p-W-{\textstyle {\frac {i}{2}}}\,\Gamma_p}\,
\label{eq:2.2.4}
\end{equation}
with $r_p$, the complex residue at the pole, given by
\begin{equation}
\left. {\rm {Res}}\, T(W)\right|_{W=W_p} = -\mid r_p \mid \, e^{i\theta_p}\,.
\end{equation}
For a single (elastic) channel, the comparison with Eq.~(\ref{eq:2.1.2.a})
yields the residue. The more general ansatz Eq.~(\ref{eq:2.2.4}) leads to the
speed
\begin{equation}
\mathrm{SP} (W) = \frac{|r_p|}{(W-M_p)^2+{\textstyle {\frac
{1}{4}}}\Gamma_p^2}\,, \label{eq:2.2.5}
\end{equation}
which is fitted to speed data obtained from the partial wave amplitudes in the
vicinity of the maximum. The pole parameters $M_p,\Gamma_p$ and $|r_p|$ are
then obtained from the best fit to the selected speed points. Finally, the
phase $\theta_p$ of the residue is obtained from the phase of $dT/dW$,
\begin{equation}
\mathrm{tan}\,\theta_p=\left.\frac{\mathrm{Im}\,(dT/dW)}{\mathrm{Re}\,(dT/dW)}\right|_{W=M_p}\,.
\label{eq:2.2.6}
\end{equation}
\subsection{Regularization method}
\label{subsec2c}
As discussed before, the speed plot technique is successful  if the background
under the resonance is approximately constant. It fails if the background
changes over the resonance region. The {\emph {regularization method}} extends
the idea behind the speed plot by construction of higher derivatives,
$T^{(N)}\equiv d^N\,T/dT^N$, $N\geq 1$~\cite{Ceci:2008zza}. Let there be an
analytic function $T(z)$ of a complex variable $z$ with a first-order pole at
some complex point $\mu=x+iy$. This function can be any of the $T$-matrix
elements, and the variable $z$ is identified with the c.m. energy $W$ in order
to compare with the speed plot technique. The described function takes the form
\begin{equation}
T(z)=\underbrace{\,\,\,\frac{r}{\mu-z}\,\,\,}_{\mathrm{resonant\,
part}}+\,\,\underbrace{\left(T(z)-\frac{r}{\mu-z}\right)}_{\mathrm{smooth\,background}} \,,
\label{eq:2.3.1}
\end{equation}
where $\mu$ and $r$ are the position and residue of the pole.  Of course, the
experiment can determine the $T$-matrix elements only for real values of $W$.
In order to continue $T(W)$ into the complex energy plane and to search for the
pole position, we construct a regular function $f$ by multiplying $T$ with the
factor $\mu-z$,
\begin{equation}
f(z)=(\mu-z)\,T(z)\, ,
\label{eq:2.3.2}
\end{equation}
with $f(\mu)=r$. In the neighborhood of the pole, the function $f$ can be
expanded in  a Taylor series. Because the scattering matrix can be accessed for
real arguments only, we construct $f(\mu)$  from the derivatives of $f$ taken
along the real axis,
\begin{equation}
r=f(\mu)=\sum_{n=0}^N \frac{f^{(n)}(W)}{n!}(\mu-W)^n+R_{N}(W,\mu).
\label{eq:2.3.3}
\end{equation}
This expansion is explicitly written to order $N$, and $R_{N}(W, \mu)$ stands
for the higher orders. The derivatives $f^{(n)}(W)$ can be turned in
derivatives of the $T$ matrix by use of Eq.~(\ref{eq:2.3.2}), and mathematical
induction leads to the following equation:
\begin{equation}
f^{(n)}(W)=(\mu-W)\,T^{(n)}(W)-n\,\,T^{(n-1)}(W).
\label{eq:2.3.4}
\end{equation}
Insertion of these derivatives into Eq.~(\ref{eq:2.3.2}) cancels all the
terms in the sum except for the last one,
\begin{equation}
r=\frac{T^{(N)}(W)}{N!}(\mu-W)^{(N+1)}+R_{N}(W,\mu)\,.
\label{eq:2.3.5}
\end{equation}
In the neighborhood of the pole, the remainder $R_{N}$
should decrease with increasing $N$. Assuming that the higher derivatives
can be neglected for a sufficiently large value of $N$ and taking the absolute values
of both sides, we obtain an approximation of the pole parameters at ${\mathcal {O}}(N)$,
\begin{equation}
\left|r_N \right|=\frac{\left|T^{(N)}(W)\right|}{N!}\left|\mu_N-W\right|^{(N+1)}.
\label{eq:2.3.6}
\end{equation}
On condition that the Taylor series converges and in the limit $N\rightarrow
\infty$, $r_N$ and $\mu_N$ should approach the values $r$ and $\mu$,
respectively. In the next step,  we (i) write the pole position as a general
complex number, $\mu=a+i\,b$, (ii) raise both sides of Eq.~(\ref{eq:2.3.6}) to
the power of $2/(N+1)$, and collect the information about the $T$-matrix and
the pole position on the right and left side, respectively. The result is a
parabolic equation in $W$,
\begin{equation}
\frac{(a_N-W)^2+b_N^2}{\sqrt[N+1]{|r_N|^2}}  = \sqrt[N+1]{\frac{\left(N!\right)^2}{\left|T^{(N)}(W)\right|^2}}\;.
\label{eq:2.3.7}
\end{equation}
This equation relates the pole position ($a=M_p,\, b=-{\textstyle {\frac
{i}{2}}}\,\Gamma_p$) and the absolute value of the residue, $|r|$, to the
T-matrix values on  the real axis, as obtained from a model or an
energy-dependent partial-wave analysis of the data. Finally, the phase of the
residue is determined by
\begin{equation}
{\mathrm{tan}}\, \theta_N = \left.\frac{\mathrm{Im}\, T^{(N)}(W)}{\mathrm{Re}\,T^{(N)}(W)}\right|_{W=M_p}\,.
\end{equation}
The comparison of the above equations with the results of Sect.~\ref{subsec2b}
shows that the speed plot is identical to the  regularization method for $N=1$.
\\

The further procedure is as follows: (i) Construct the $N^{th}$ derivative of
the $T$-matrix element  and the right-hand side of Eq.~(\ref{eq:2.3.7}). Note
that the pole parameters are uniquely determined by the {\emph{exact}}
knowledge of $T(W)$ in only {\emph{ three points}}. However, the problem is to
choose the right points. If the distance between the points gets too large, the
influence of other singularities may increase. If the points are too close,
numerical problems may occur. (ii) Solve Eq.~(\ref{eq:2.3.7}) for the pole
parameters by either choosing various three-point sets to evaluate the
right-hand side and perform a statistic analysis of the results or fitting the
right-hand side of the equation to a three-parameter parabolic function. In our
approach we have chosen the latter option.\\

In closing this section we note that the regularization method does  not depend
on any particular functional form of the $T$ matrix. However, we have to assume
that (a) the $N^{th}$ derivative can be constructed with a sufficient precision
and (b) the pole position lies within the circle of convergence for the Taylor
expansion, that is, no further singularities should intrude into the region
between the pole and the related resonance region on the real $W$ axis.
\subsection{Poles from analytic continuation}
\label{subsec2d}
The most accurate way to determine pole positions and residues is certainly
obtained by analytic continuation into the complex region. Because resonance
poles can not appear on the physical sheet, we have to take a careful look at
the structure of different Riemann sheets opening at all branch points for
particle production in a coupled-channels model. The most important particle
thresholds in our energy region below 2~GeV are $\pi\pi N$, $\pi\Delta$, $\eta
N$, and $\rho N$ with branch points at $1178$~MeV, $(1350-50i)$~MeV,
$1486$~MeV, and $(1713-75i)$~MeV, in order. In the dynamical DMT model we have
included the $\pi N$ and $\pi\pi N$ channels in all the partial waves and the
$\eta N$ channel in the $S_{11}$ partial wave. However, the $\pi\pi N$ channel
is treated in a phenomenological way as will be described in the following
section. Because the particular ansatz for the two-pion width,
Eq.~(\ref{eq:3.2.6}), contains only even powers of $q_{2\pi}$, the model does
not give additional branch points for the two-pion channel. This leads to a
relatively simple sheet structure, and  we can easily reach the poles on the
most important second Riemann sheet (first unphysical sheet). Technically, we
first map the relevant region on this sheet by contour plots and search for the
accurate pole position by standard root finding routines applied to the
function
\begin{equation}
h(z_p)=\frac{1}{|T(z_p)|}=0\,.
\end{equation}
Next we obtain the residue by approaching the pole along
different paths in the complex plane,
\begin{equation}
\left. {\rm {Res}}\,T(z)\right|_{z_p} = \lim_{z\to z_p}(z-z_p)\,T(z)\,.
\end{equation}
A word of caution is in order. If the fitted form factor parameters, e.g.,
$\Lambda_\alpha$ of Eq.~(\ref{eq:3.1.11}) or $X_R$ of Eq.~(\ref{eq:3.2.6})
become smaller than about $500$~MeV, additional poles can appear in the region
where the resonance poles are expected. In order to avoid such spurious
singularities, it is very important to map out the structure of the $T$ matrix
very precisely. Also the speed-plot and regularization methods are very helpful
to distinguish between resonance and spurious poles, because the latter ones
usually affect $T$ matrix at real $W$ in a similar way as a broad background does.
\\

\section{The Dubna-Mainz-Taipei meson-exchange model}
\label{sec3}
The Dubna-Mainz-Taipei (DMT) $\pi N$ meson-exchange model was developed  on the
basis of the Taipei-Argonne $\pi N$ meson-exchange
model~\cite{Lee:1991dd,Hung:1994hg,Hung:2001pz} which describes pion-nucleon
scattering up to 400~MeV pion laboratory energy. The DMT model was extended to
c.m. energies $W=2.0$~GeV by inclusion of higher resonances and the $\eta N$
channel~\cite{Chen:2002mn,Chen:2007cy}. The model gives an excellent fit to
both $\pi N$ phase shifts and inelasticity parameters in all the channels up to the
$F$ waves and energies of 2~GeV, except for the $F_{17}$ partial wave. The DMT
$\pi N$ model is also a main ingredient of the Dubna-Mainz-Taipei dynamical
model describing the photo- and electroproduction of
pions~\cite{Kamalov:1999hs} up to 2~GeV. In particular, this model gives an
excellent agreement with pion production data from threshold to the first
resonance region~\cite{Kamalov:2000en,Kamalov:2001qg}. In this section, we
briefly outline the ingredients of the Taipei-Argonne $\pi N$ model and then
describe some relevant features of the DMT meson-exchange coupled-channels
model.
\subsection{Taipei-Argonne meson-exchange $\pi N$ model}
\label{subsec3a}
The Taipei-Argonne model describes the elastic scattering of pions and
nucleons. It is based on a three-dimensional reduction of the Bethe-Salpeter
equation for an effective Lagrangian involving $\pi$, $N$, $\Delta$, $\rho$,
and $\sigma$ fields. Let us first describe some kinematics for $\pi N$
scattering,
\begin{equation}
\pi (q) + N(p) \rightarrow \pi (q') + N(p')\,,
\label{eq:3.1.1}
\end{equation}
where $q, p, q'$, and $p'$ are the four-momenta of the
respective particles. The total and relative
four-momenta are $P = p+q$ and $k  = p \,\eta_{\pi}(s)-q \,\eta_N(s)$,
respectively, where $s = P^2 = W^2$ is the Mandelstam variable. The
dimensionless variables $\eta_{\pi}(s)$ and $\eta_N(s)$ represent the
freedom in choosing a three-dimensional reduction, they are
constrained by the condition $\eta_N + \eta_{\pi} = 1$. For further
details we refer to Ref.~\cite{Hung:2001pz}.\\

The Bethe-Salpeter equation for $\pi N$ scattering takes the form
\begin{equation}
T_{\pi N} = B_{\pi N} + B_{\pi N}G_0 T_{\pi N}\,,
\label{eq:3.1.2}
\end{equation}
where $B_{\pi N}$ is the sum of all irreducible two-particle Feynman
amplitudes and $G_0$ the free relativistic pion-nucleon propagator.
Equation~(\ref{eq:3.1.2}) can be cast into the form
\begin{equation}
 T_{\pi N} = \hat B_{\pi N} + \hat B_{\pi N}\hat G_0T_{\pi N}\,,
 \label{eq:3.1.3}
\end{equation}
with
\begin{equation}
\hat B_{\pi N} = B_{\pi N} + B_{\pi N}(G_0-\hat G_0)\hat B_{\pi N}\,,
\label{eq:3.1.4}
\end{equation}
where $\hat G_0(k;P)$ is an appropriate propagator to obtain a
three-dimensional reduction of Eq.~(\ref{eq:3.1.2}). This propagator is chosen
to maintain the two-body unitarity by reproducing the $\pi N$ elastic cut.
However, there is still a wide range of possible propagators satisfying this
constraint. A standard choice is the Cooper-Jennings reduction
scheme~\cite{Cooper:1988ap},
\begin{eqnarray}
\hat G_0(k;P) & = & \frac{1}{(2\pi)^3}\int \frac{ds'}{ s-s'} f(s,s')\,
[\alpha (s,s') P \hspace{-0.10in} / +k
\hspace{-0.08in} / +m_N]  \nonumber  \\
& & \times\, \delta^{(+)}([\eta_N(s')P' + k]^2 - m_{N}^2)\,  \delta^{(+)}
([\eta_{\pi} (s')P'- k]^2 -  m_{\pi}^2)\,,
\label{eq:3.1.5}
\end{eqnarray}
with $P' = \sqrt{s'/s}\,P$ and the superscript (+) of the  $\delta$-functions
standing for the positive energy part of the propagator. The variables $f$ and
$\alpha$ are dimensionless variables containing the freedom of reduction, they
are constrained by the conditions  $ f(s,s) = 1$ and $\alpha(s,s) = \eta_N(s)$
to ensure the reproduction of the elastic cut. Furthermore, Cooper and Jennings
define~\cite{Cooper:1988ap}
\begin{equation}
\alpha(s,s')  =  \eta_N(s)\,, \quad f(s,s')=
\frac{4\sqrt{ss'}\varepsilon_N(s')\varepsilon_\pi(s')}{ss'-(m_N^2-m_{\pi}^2)^2}\,,
\label{eq:3.1.6}
\end{equation}
which leads to the following expression for $\hat G_0$ in the c.m. frame:
\begin{equation}
\hat G_0(k;s)=\frac{1}{(2\pi)^3}
\frac{\delta[k_0-\hat{\eta}(s_{\vec{k}}, \vec{k})]}
{\sqrt{s}-\sqrt{s_{\vec{k}}}} \frac{2 \sqrt{s_{\vec{k}}}} {\sqrt{s}+
\sqrt{s_{\vec{k}}}} f(s,s_{\vec{k}}) \frac{\alpha(s,s_{\vec{k}})
\gamma_0 \sqrt{s} +k \hspace{-0.08in} / + m_N} {4 E_N(\vec{k})
E_\pi(\vec{k})}\,,
\label{eq:3.1.7}
\end{equation}
where  $E_N(\vec{k})$ and  $E_\pi(\vec{k})$ are the nucleon and pion
energies for the three-momentum $\vec k$, $\sqrt{ s_{\vec{k}} }  =
E_N(\vec{k})+E_\pi(\vec{k})$ is the total energy in the c.m. frame,
and $\hat{\eta}(s,\vec{k}) = E_N(\vec{k})-\eta_N(s_{\vec{k}})
\sqrt{s_{\vec{k}}}\, .$ With these relations we obtain the
following $\pi N$ scattering equation:
\begin{equation}
t(\vec{k'},\vec{k};W) = v(\vec{k'},\vec{k};W) + \int d
\vec{k''}v(\vec{k'},\vec{k''};W) g_0(\vec k'';W)
t(\vec{k''},\vec{k};W)\, . \label{eq:3.1.8}
\end{equation}
The explicit relations between the quantities defined by Eqs.~(\ref{eq:3.1.2})
and (\ref{eq:3.1.8}) are given by
\begin{eqnarray}
t(\vec{k'},\vec{k};W) & = &\int dk'_0 dk_0\delta(k'_0-\hat{\eta}')
T(k',k;W)
\delta(k_0-\hat{\eta}),\label{eq:3.1.9.a}  \\
v(\vec{k'},\vec{k};W) & = &\int dk'_0 dk_0\delta(k'_0-\hat{\eta}')
B(k',k;W)
\delta(k_0-\hat{\eta}), \label{eq:3.1.9.b} \\
g_0(\vec k;W) & = & \int dk_0 \hat G_0(k;W)\,,\label{eq:3.1.9.c}
\end{eqnarray}
with $\hat{\eta}'=  \hat{\eta}(s_{\vec{k'}}, \vec{k'})$ and
$\hat{\eta}= \hat{\eta}(s_{\vec{k}}, \vec{k})$.\\

The effective Lagrangian involving the $\pi, N, \sigma, \rho,$
and $\Delta(1232)$ fields takes the form
\begin{eqnarray}
{\cal L}_I & = & \frac{f_{\pi NN}^{(0)}}{m_\pi} \bar N \gamma_5
\gamma_\mu \vec{\tau} \cdot \partial^\mu \vec{\pi} N
- g^{(s)}_{\sigma\pi\pi} m_{\pi} \sigma (\vec{\pi}\cdot\vec{\pi})
- \frac {g^{(v)}_{\sigma\pi\pi}}{2  m_{\pi}} \sigma
(\partial^{\mu}\vec{\pi}\cdot\partial_{\mu}\vec{\pi})
-g_{\sigma NN}\bar{N}\sigma N \nonumber \\
& & -g_{\rho NN} \bar{N} \{ \gamma_\mu \vec{\rho}\,{}^\mu +
\frac{\kappa_V^\rho}{4m_N} \sigma_{\mu\nu} (\partial ^\mu
\vec{\rho}\,{}^\nu -
\partial^\nu \vec{\rho}\,{}^\mu) \} \cdot \frac{1}{2}\vec{\tau} N  \nonumber \\
& & -g_{\rho\pi\pi} \vec{\rho}\,{}^{\mu} \cdot (\vec{\pi} \times
\partial_{\mu} \vec{\pi}) -
\frac{g_{\rho\pi\pi}}{4m_{\rho}^2}(\delta - 1)(\partial^\mu
\vec{\rho}\,{}^\nu -\partial^\nu \vec{\rho}\,{}^\mu) \cdot
(\partial_\mu \vec{\pi} \times\partial_\nu \vec{\pi})  \nonumber \\
& & + \frac{g_{\pi N\Delta}}{m_\pi}  \bar\Delta_\mu [g^{\mu\nu}
-(Z+\frac{1}{2})\gamma^\mu\gamma^\nu] \vec{T}_{\Delta N}N \cdot
\partial_{\nu} \vec \pi\,,
\label{eq:3.1.10}
\end{eqnarray}
with  $\Delta_{\mu}$ the Rarita-Schwinger field operator for the
$\Delta$ resonance and $\vec T_{\Delta N}$ the isospin transition
operator between nucleon and $\Delta$. The driving term
$B_{\pi N}$ in Eq.~(\ref{eq:3.1.2}) is approximated by the tree
diagrams of the interaction Lagrangian of Eq.~(\ref{eq:3.1.10}),
the direct and crossed $N$ and $\Delta$ diagrams as well
as the t-channel $\sigma$- and $\rho$-exchange contributions.\\

Furthermore, the driving term $v(\vec{k},\vec{k}';W)$ of
Eq.~(\ref{eq:3.1.9.b}) is regularized by covariant form factors
associated with each leg of the vertices,
\begin{equation}
F_{\alpha}(p^2_{\alpha})=\left[\frac {n\Lambda_{\alpha}^4}{n\Lambda_{\alpha}^4 + (m_{\alpha}^2 - p_{\alpha}^2)^2}\right]^{n}\,,
\label{eq:3.1.11}
\end{equation}
with $p_{\alpha}$ the four-momentum and $m_{\alpha}$ the mass of particle $\alpha$. In the present
work we choose the value $n = 10$ as used by Ref.~\cite{Chen:2002mn}.\\

As in the work of Afnan and collaborators~\cite{Morioka:1981vb,Pearce:1986du},
the $P_{11}$ phase shift is constrained by imposing the nucleon pole condition. This
treatment leads to a proper renormalization of both nucleon mass and $\pi NN$
coupling constant. It also yields the necessary cancelation between the
repulsive nucleon pole contribution and the attractive background, such that a
reasonable fit to the $\pi N$ phase shifts in the $P_{11}$ channel can be
achieved.\\

The following parameters were allowed to vary in the fit procedure: the
products $g_{\sigma NN}g^{(s)}_{\sigma\pi\pi}, \, g_{\sigma
NN}g^{(v)}_{\sigma\pi\pi}$, and $g_{\rho NN}g_{\rho\pi\pi}$ as well as $\delta$
for the t-channel $\sigma$ and $\rho$ exchanges, $m_{\Delta}^{(0)},\,
g^{(0)}_{\pi N\Delta}$, and $Z$ for the $\Delta$ mechanism, and the cut-off
parameters $\Lambda_{\alpha}$ of the form factors given by
Eq.~(\ref{eq:3.1.11}). The experimental $\pi N$ phase shifts were well
described up to pion laboratory energies of 400~MeV. The resulting parameters
and predicted phase shifts can be found in Ref. \cite{Hung:2001pz}.
\\
\subsection{DMT meson-exchange model including $\eta N$ channel and higher resonances}
\label{subsec3b}
As the energy gets larger, heavier mass channels like $\sigma N,\, \eta N,
\,\pi\Delta$, and $\rho N$ as well as a non-resonant continuum of $\pi \pi N$
states become increasingly important, and at the same time more and more
nucleon resonances appear as intermediate states. For each contributing
resonance $R$, the Hilbert space is enlarged by a bare resonance $R$ which
acquires a width by its coupling to the $\pi N$ and $\eta N$ channels through
the Lagrangian
\begin{equation}
{\mathcal{L_I}}= ig^{(0)}_{\pi NR}\bar R \tau N\cdot \pi +
ig^{(0)}_{\eta NR}\bar R N\eta + {\mathrm{h.c.}} \,,
\label{eq:3.2.1}
\end{equation}
where $N, R, \pi,$ and $\eta$ denote the field operators for the
nucleon, bare resonance $R$, pion, and eta meson, respectively. The
full $t$-matrix can be written as a system of coupled equations,
\begin{equation}
t_{ij}(W)= v_{ij}(W)+\sum_k  v_{ik}(W)\,g_k(W)\, t_{kj}(W)\,,
\label{eq:3.2.2}
\end{equation}
with $i$ and $j$ denoting the $\pi$ and $\eta$ channels and $W$ is
the total c.m. energy.\\

In general, the potential $ v_{ij}$ is the sum of non-resonant
$(v^B_{ij})$ and bare resonance $(v^R_{ij})$ terms,
\begin{equation}
v_{ij}(W)=  v^B_{ij}(W)+ v^R_{ij}(W)\,. \label{eq:3.2.3}
\end{equation}
The non-resonant term $v^B_{\pi\pi}$ for the $\pi N$ elastic channel,
as defined in Sec.~\ref{subsec3a}, contains
contributions from $s$- and $u$-channel Born terms as well as
$t$-channel contributions with $\omega$, $\rho$, and $\sigma$
exchange. The parameters in $v^B_{\pi\pi}$ are fixed by an
analysis of the pion scattering phase shifts for the $S$ and
$P$ waves at energies $W<1300$~MeV~\cite{Hung:2001pz}. In the channels
involving the $\eta$, the potential $v^B_{i\eta}$ is assumed to vanish
because of the small $\eta NN$ coupling~\cite{Tiator:1994et}.\\

The bare resonance contribution arises from the excitation and
decay of the resonance $R$,
\begin{equation}
v^R_{ij}(W)=\frac {h_{i R}^{(0)\dagger} h_{j
R}^{(0)}}{W-M_R^{(0)}}\,, \label{eq:3.2.4}
\end{equation}
where $M_R^{(0)}$ and $h_{i R}^{(0)}$ denote the bare mass for the
resonance $R$ and the bare vertex operator $R\rightarrow \pi/\eta +
N$, respectively. The matrix elements of the potential $v^R_{ij}(W)$
is symbolically expressed by
\begin{equation}
v^R_{ij}(q,q';W)=\frac{f_i(\tilde
{\Lambda}_i,q;W)\,g_i^{(0)}\,g_j^{(0)}\,
f_j(\tilde{\Lambda}_j,q';W)}{W-M_R^{(0)}+
\frac{i}{2}\Gamma_R^{2\pi}(W)} \,, \label{eq:3.2.5}
\end{equation}
where $q$ and $q'$ are the pion (or eta) momenta in the initial and
final states, and $g_{i/j}^{(0)}$ denotes the resonance vertex
couplings. As in Ref.~\cite{Hung:2001pz}, we associate a covariant
form factor with each leg of the vertices.  Therefore, each vertex
$f_i$ in Eq.~(\ref{eq:3.2.5}) contains three form factors as given by
Eq.~(\ref{eq:3.1.11}), and $\tilde\Lambda_i$ stands for a triple of
cut-offs, ($\Lambda_N,\Lambda_R,\Lambda_\pi$). In
Eq.~(\ref{eq:3.2.5}) we have also included a phenomenological term
$\Gamma_R^{2\pi}(W)$ in the resonance propagator to account for the
$\pi\pi N$ decay channel. Therefore, our ``bare'' resonance
propagator already contains a phenomenological ``dressing'' effect
due to the coupling to the $\pi\pi N$ channel. With this
prescription we assume that any further non-resonant coupling to the
$\pi\pi N$ channel can be neglected. Following
Refs.~\cite{Lvov:1996xd,Drechsel:1998hk} we parameterize the
two-pion width by
\begin{equation}
\Gamma_R^{2\pi}(W)=\Gamma^{2\pi(0)}_{R}\left(\frac{q_{2\pi}}{q_0}\right)^{2\ell+4}
\left(\frac{X_R^2+q^2_0}{X_R^2+q^2_{2\pi}}\right)^{\ell+2}\,,
\label{eq:3.2.6}
\end{equation}
where $\ell$ is the pion orbital momentum and $q_{2\pi}=q_{2\pi}(W)$
the momentum of the compound two-pion system. Furthermore, the two-pion width and
two-pion momentum at resonance are denoted by $\Gamma^{2\pi(0)}_{R}$  and
$q_0$, respectively. We note that this form accounts for the
correct energy behavior of the phase space near the three-body
threshold~\cite{Lvov:1996xd}. In our present work,
$\Gamma_{R}^{2\pi(0)}$ and $X_R$ are considered as free parameters.
Therefore, each resonance is generally described by 6 free
parameters, the bare mass $M^{(0)}_R$, the decay width
$\Gamma^{2\pi(0)}_{R}$, two bare coupling constants $g^{(0)}_i$ and
$g^{(0)}_j$, and two cut-off parameters $\Lambda_R$ and $X_R$. The
generalization of the coupled-channel model to the case of $N$
resonances with the same quantum numbers is then given by
\begin{equation}
v^R_{ij}(q,q';W)=\sum_{n=1}^{N} v^{R_n}_{ij}(q,q';W)\,,
\label{eq:3.2.7}
\end{equation}
with 6 free parameters for each resonance.\\

The solutions of the coupled-channel equations of Eq.~(\ref{eq:3.2.2}), with
potentials given in Eqs.~(\ref{eq:3.2.3}-\ref{eq:3.2.6}), were fitted to the
experimental $\pi N$ phase shifts and inelasticities by variation of the bare
resonance parameters.  The fit gave a good agreement with the data for all
channels up to the $F$ waves and energies below 2~GeV, except for the partial
wave $F_{17}$~\cite{Chen:2007cy}. The predictions of the DMT model for the
resonance parameters are  presented in the next section.
\section{Results and discussion}
In this section we present the nucleon resonance parameters as derived from the
DMT model. The listed resonances fulfill the following criteria: (i) the pole
position is restricted by $M_p\leq 2$~GeV and $\Gamma_p\leq 0.4$~GeV, (ii) the
residue is larger than about 1~MeV, and (iii) the branching ratio for the
one-pion channel is limited by $2\, r_p/\Gamma_p \geq 10$~\%. Furthermore, the
pole position obtained by the renormalization method has to be stable over a
range of derivatives (N values).

\subsection{S-waves}
As reported in Ref.~\cite{Chen:2002mn}, we need four $S_{11}$ resonances to fit
the $\pi N$ scattering amplitude in this channel, instead of the three
resonances listed by the PDG~\cite{PDG2008}. The additional resonance
$S_{11}(1878)$ was found to play an important role in pion photoproduction as
well~\cite{Chen:2002mn}, but was not seen in both the $\pi N \rightarrow\eta N$
reaction and recent measurements of $\eta$ photoproduction from the
proton~\cite{Thoma:2005awa,Elsner:2007hm}. However, in the analysis of Durand
{\emph {et al.}}~\cite{Durand:2008es,Durand:2008kw}  on eta photoproduction a
new $S_{11}$ state of mass $M=1707$~MeV and width $\Gamma=222$~MeV was needed
in order to get good agreement with the data.\\

\begin{figure}[htb]
\begin{center}
\includegraphics[width =5.3cm, angle=90]{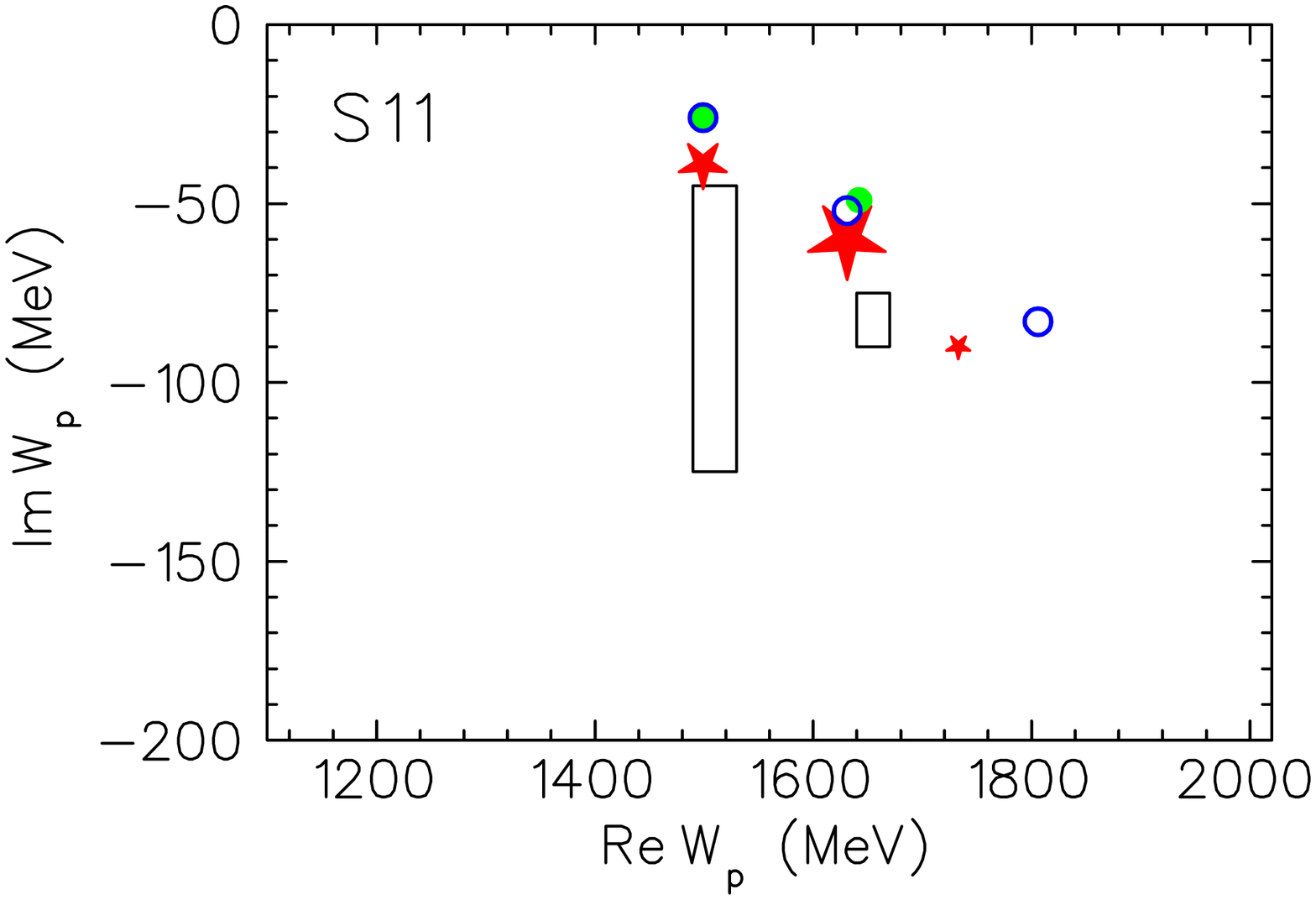}
\hspace*{0.1cm}
\includegraphics[width =5.3cm, angle=90]{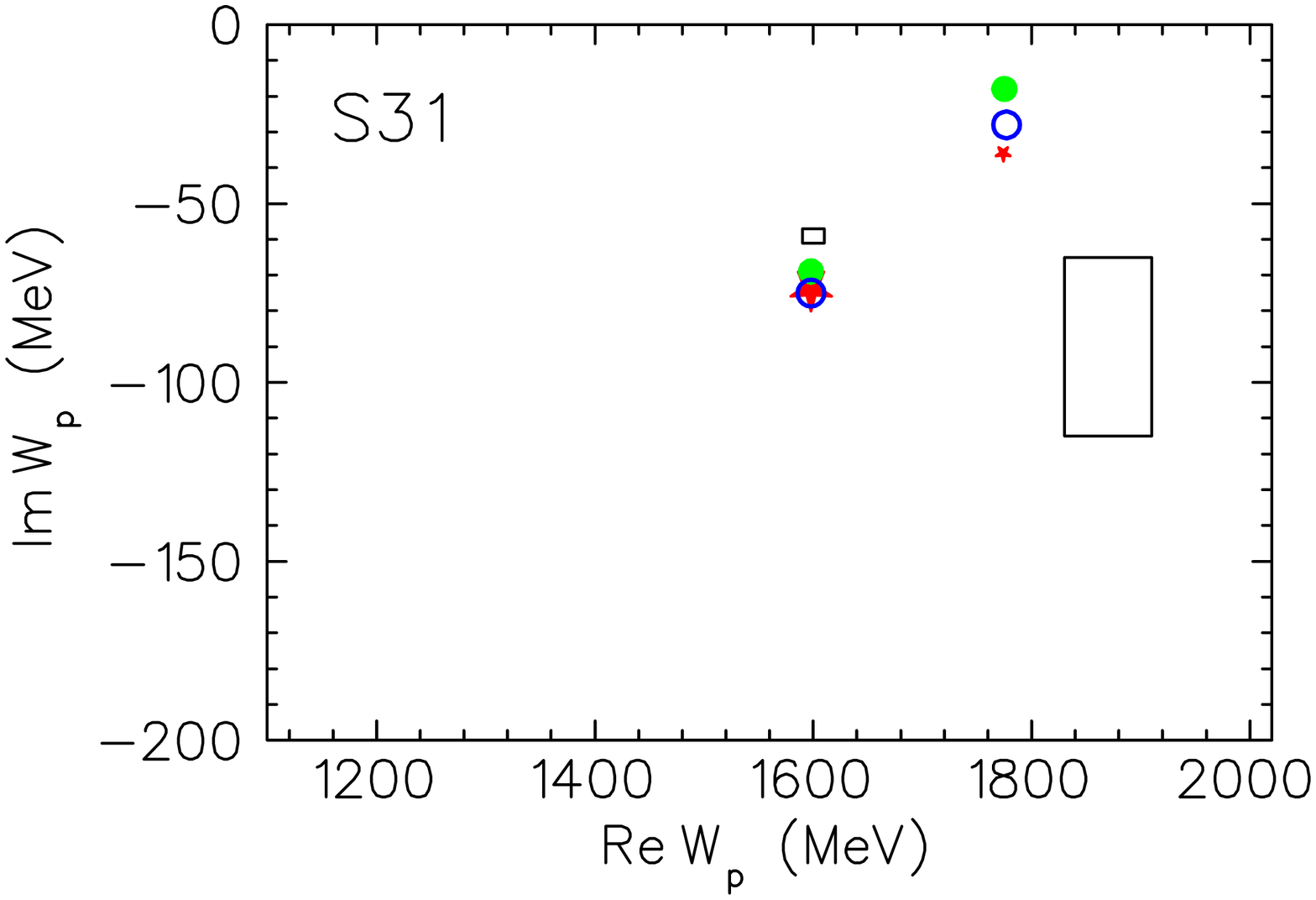}
\end{center}
\vspace*{-0.5cm} \caption{The pole positions for the $S$ waves in the complex energy plane.
The (red) stars show the results found by analytic continuation,
the (green) solid circles and the (blue) open
circles are determined by the speed-plot and the regularization
methods, respectively. The rectangular regions
show the range of the pole positions listed by the Particle Data Group
(PDG08)~\cite{PDG2008}. The size of the (red) stars is proportional
to $|r_p|/\Gamma_p$, and therefore a measure for the strength of the resonance poles.}
\label{fig:swaves}
\end{figure}

In our analysis (see Fig.~\ref{fig:swaves}, left panel) we find four poles,
three of them below 2~GeV. The exact pole positions are shown by asterisks with
a size proportional to their relative strength. Furthermore, the ranges of the
PDG pole values are displayed by open boxes. The first resonance
$S_{11}(1535)^{****}$ is well found by both speed-plot and regularization
methods. The second state, the $S_{11}(1650)^{****}$ is even better described,
with some improvement by the regularization method. The third state is very
weakly excited in the $\pi N$ interaction and can not be seen by the SP,
whereas the RM finds a state close by. A fourth state, slightly above 2~GeV and
classified by PDG as 1-star $S_{11}(2090)^{*}$ is also found by SP and RM at
close-by positions.\\

For the isospin 3/2 state we obtain three poles in agreement with PDG. The
first one, $S_{31}(1620)^{****}$ is nicely described by SP and perfectly by RM.
The second pole, $S_{31}(1900)^{**}$ is similarly well seen by both SP and RGM.
The third resonance, $S_{31}(2150)^{*}$, lies far down in the imaginary region.
Only the SP locates a state in the expected energy region, however, with only a
small fraction of the residue can be obtained.
\subsection{P-waves}
\begin{figure}
\begin{center}
\includegraphics[width =5.3cm, angle=90]{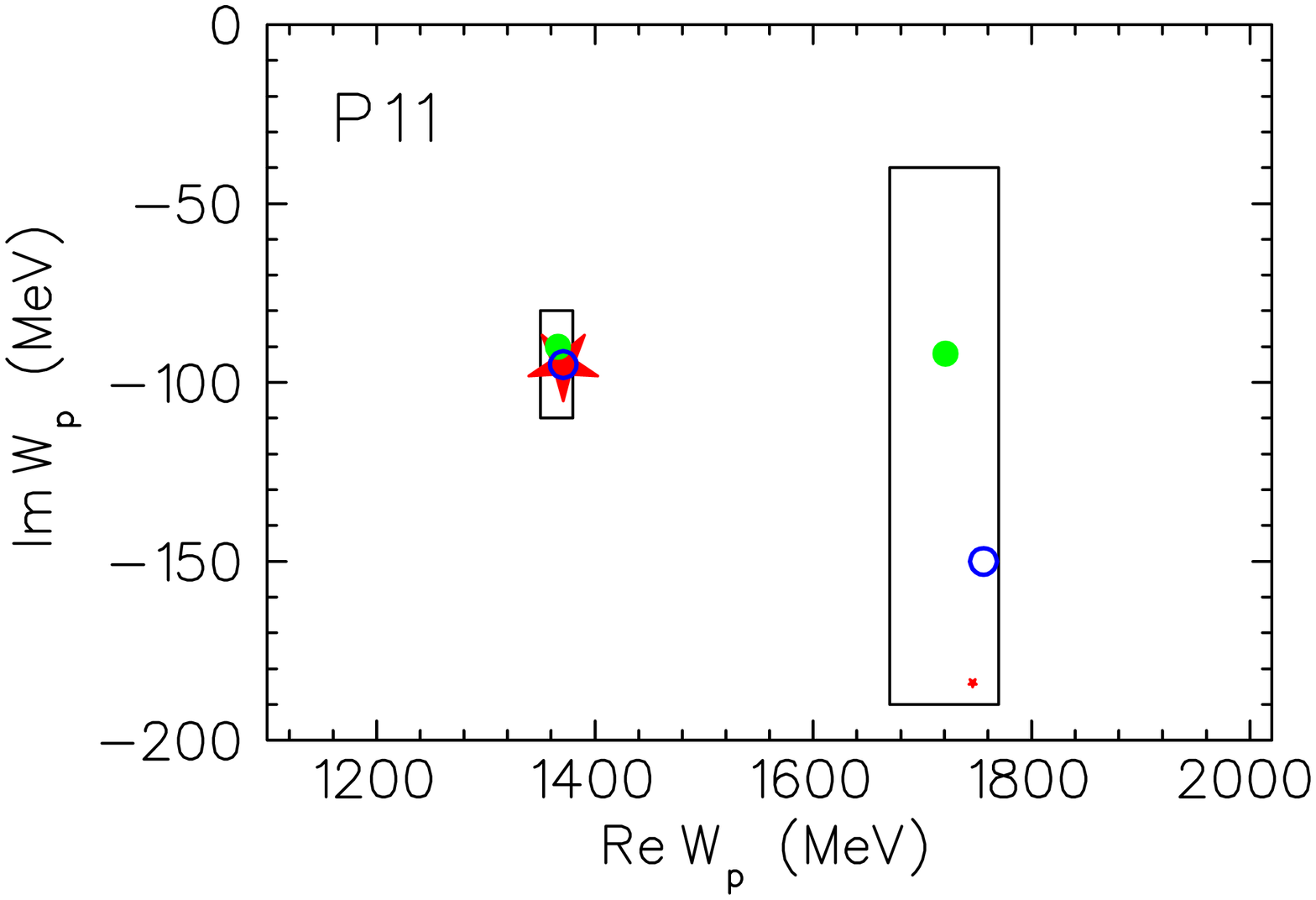}
\hspace*{0.1cm}
\includegraphics[width =5.3cm, angle=90]{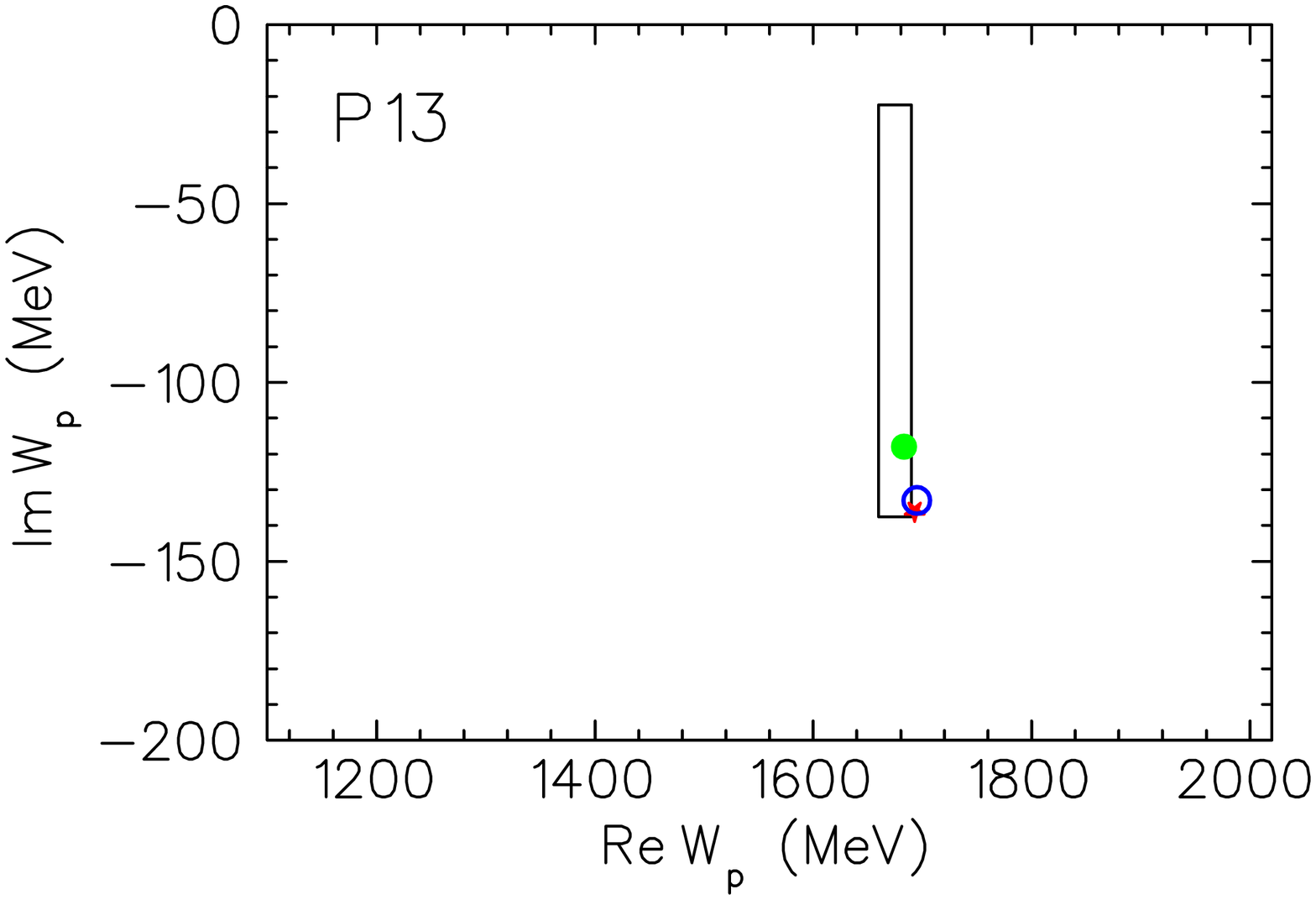}
\includegraphics[width =5.3cm, angle=90]{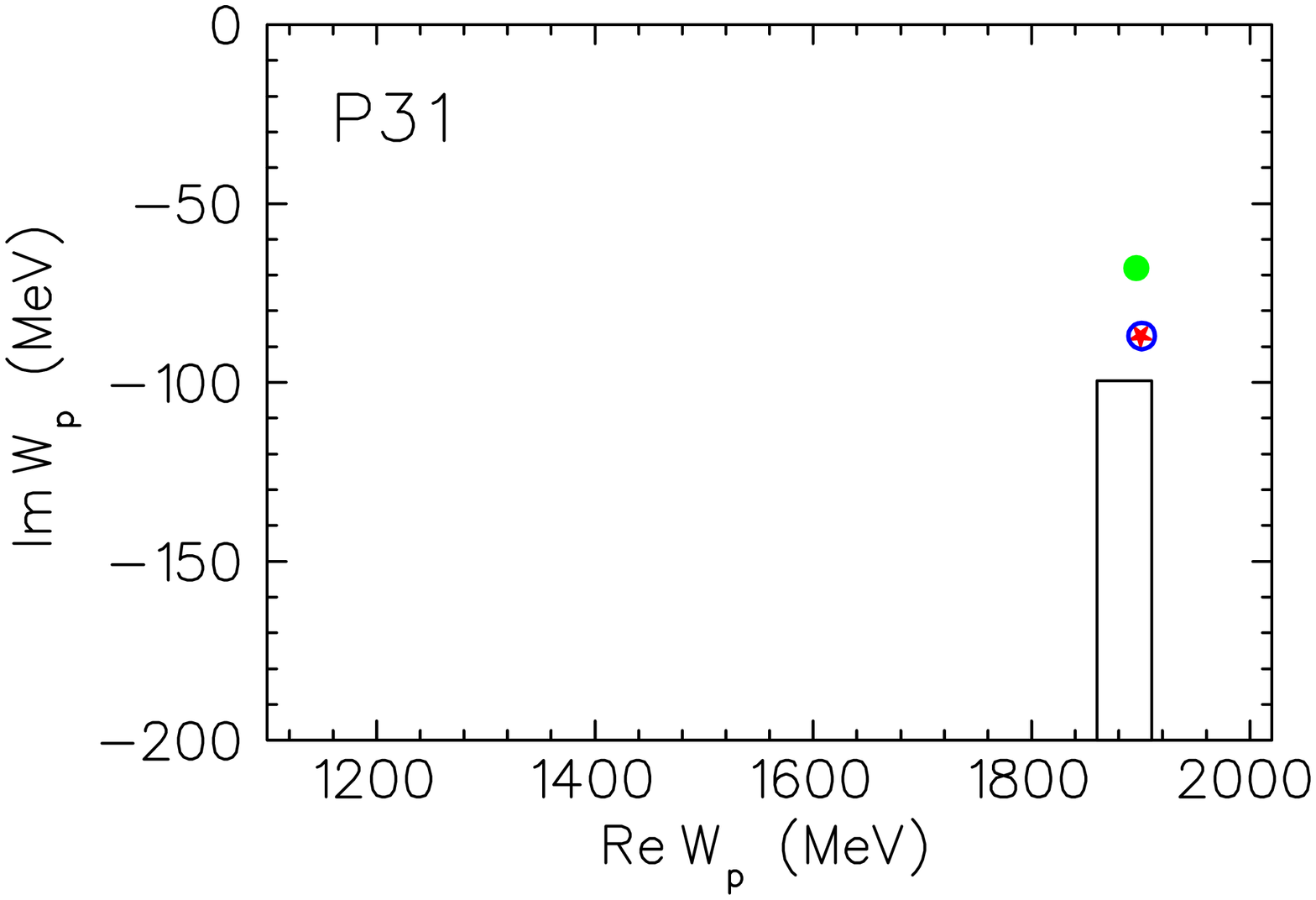}
\hspace*{0.1cm}
\includegraphics[width =5.3cm, angle=90]{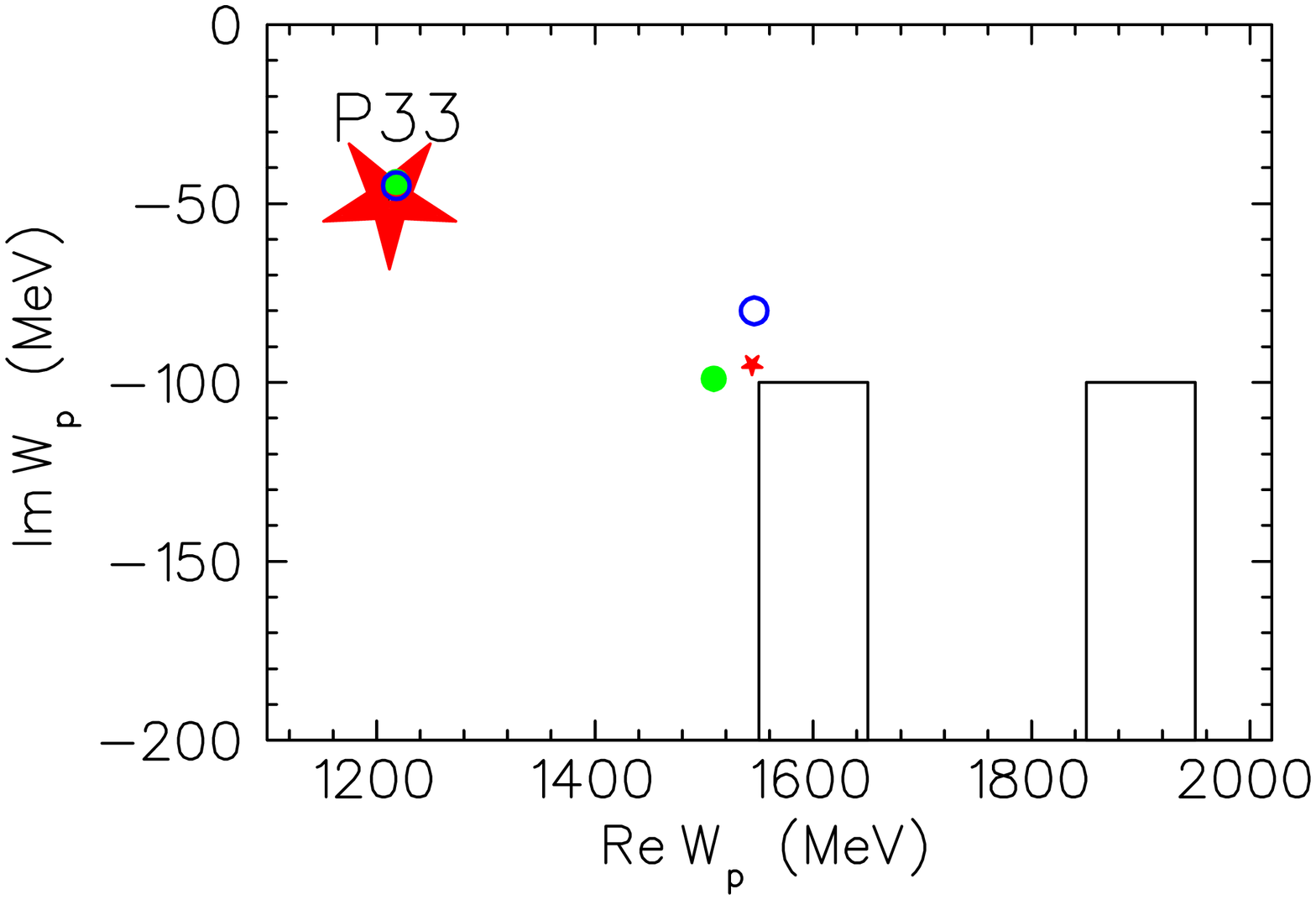}
\end{center}
\vspace*{-0.5cm} \caption{The pole positions for the $P$ waves in the complex energy plane.
The notation is the same as in Fig.~\ref{fig:swaves}.}
\label{fig:pwaves}
\end{figure}

For the $P$-wave resonances we show our results in Fig.~\ref{fig:pwaves}. In
the $P_{11}$ partial wave we get 3 poles, all of them being somehow related to
the 3 PDG states. However, our second and third pole positions lie further down
in the negative imaginary region, the third pole is even below the range shown
in the figure. The position of the first resonance, the Roper
$P_{11}(1440)^{****}$ is very well described by both SP and RM. The two higher
states have very weak signals in the $\pi N$ amplitude. Although the RM yields
a considerable improvement over the SP, it also misses the exact pole position.
\\

From the results in Table~\ref{table:Ppoles} we find that only a very small
fraction of the residue can be found with these techniques. This is a general
property for poles lying far away from the real axis. A detailed analysis of
the contour for this partial wave also shows a zero position sitting
closely above the pole position, and therefore masking the pole
if searched for along the real axis.\\

In the $P_{13}$ partial wave we also find three poles. Two of them are very
well reproduced by RM. PDG lists only two states, the $P_{13}(1720)^{****}$
with a large error bar for the imaginary part and the $P_{13}(1900)^{**}$,
however, with no pole position data given. Our results for the
$P_{13}(1720)^{****}$ lie close to the PDG values,
with the imaginary part close to the lower limit of PDG error bar.\\

The isospin-3/2 partial wave, $P_{31}$ is an exceptional case. Here the lowest
state is not the dominant one, and also the PDG lists a $P_{31}(1750)^{**}$
without pole values. We find a lowest pole near 1800~MeV with a very large
imaginary value outside of our considered range. The dominant
$P_{31}(1900)^{****}$ lies only slightly outside the PDG box
and is very well described by RM.\\

The Delta resonance, $P_{33}(1232)^{****}$, has of course the largest strength
of all the poles found in our analysis. It is very well described by both the
SP and RM techniques and coincides with PDG. Two further, rather weakly excited
$P_{33}$ states are found at higher energies, which can be related to the
states $P_{33}(1600)^{***}$ and $P_{33}(1920)^{***}$ reported by PDG. While the
analytic result for the $P_{33}(1600)^{***}$ agrees fairly well with RM, the
$P_{33}(1920)^{***}$ lies outside of the considered range.
\subsection{D-waves}
\begin{figure}
\begin{center}
\includegraphics[width =5.0cm, angle=90]{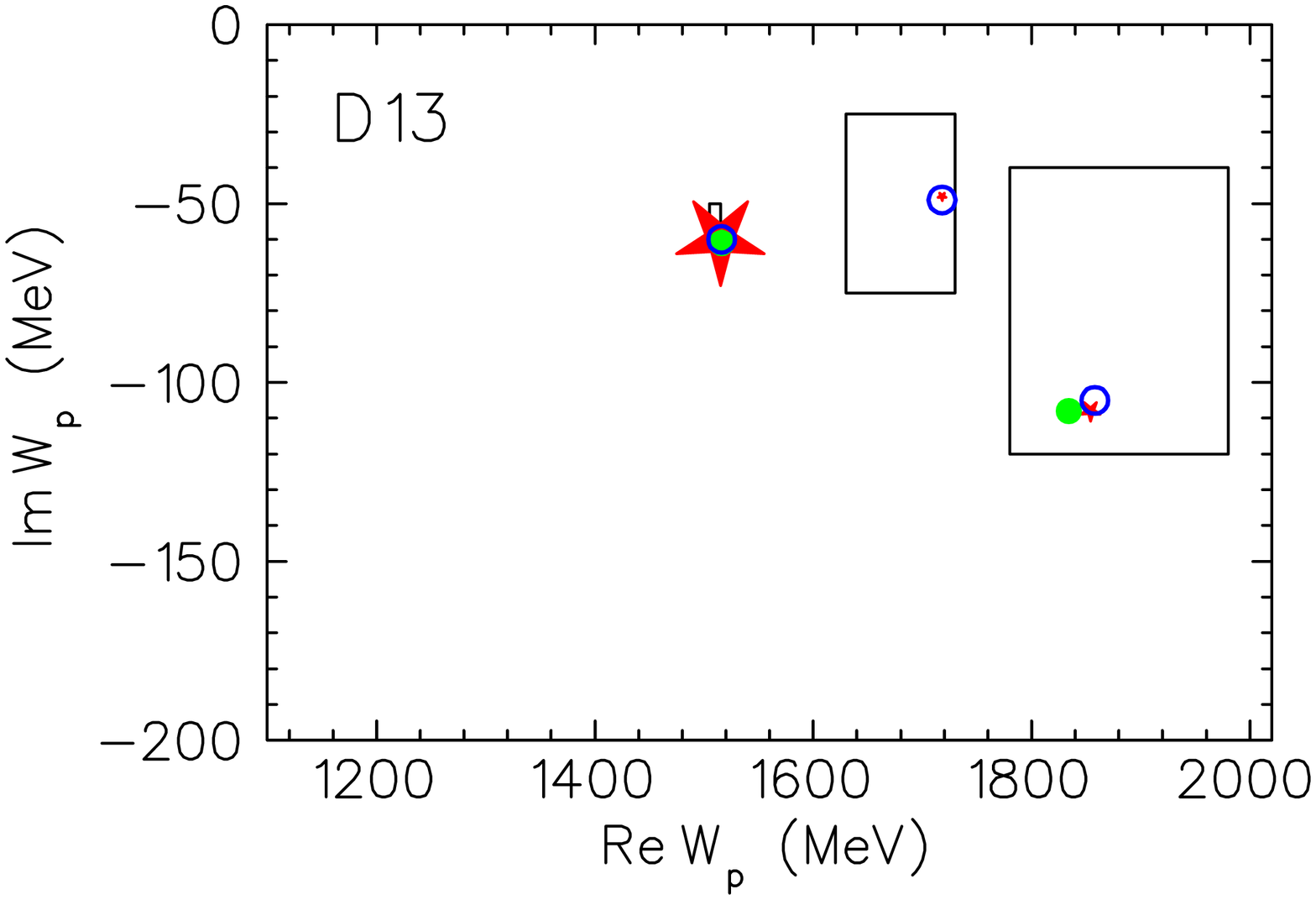}
\hspace*{0.1cm}
\includegraphics[width =5.0cm, angle=90]{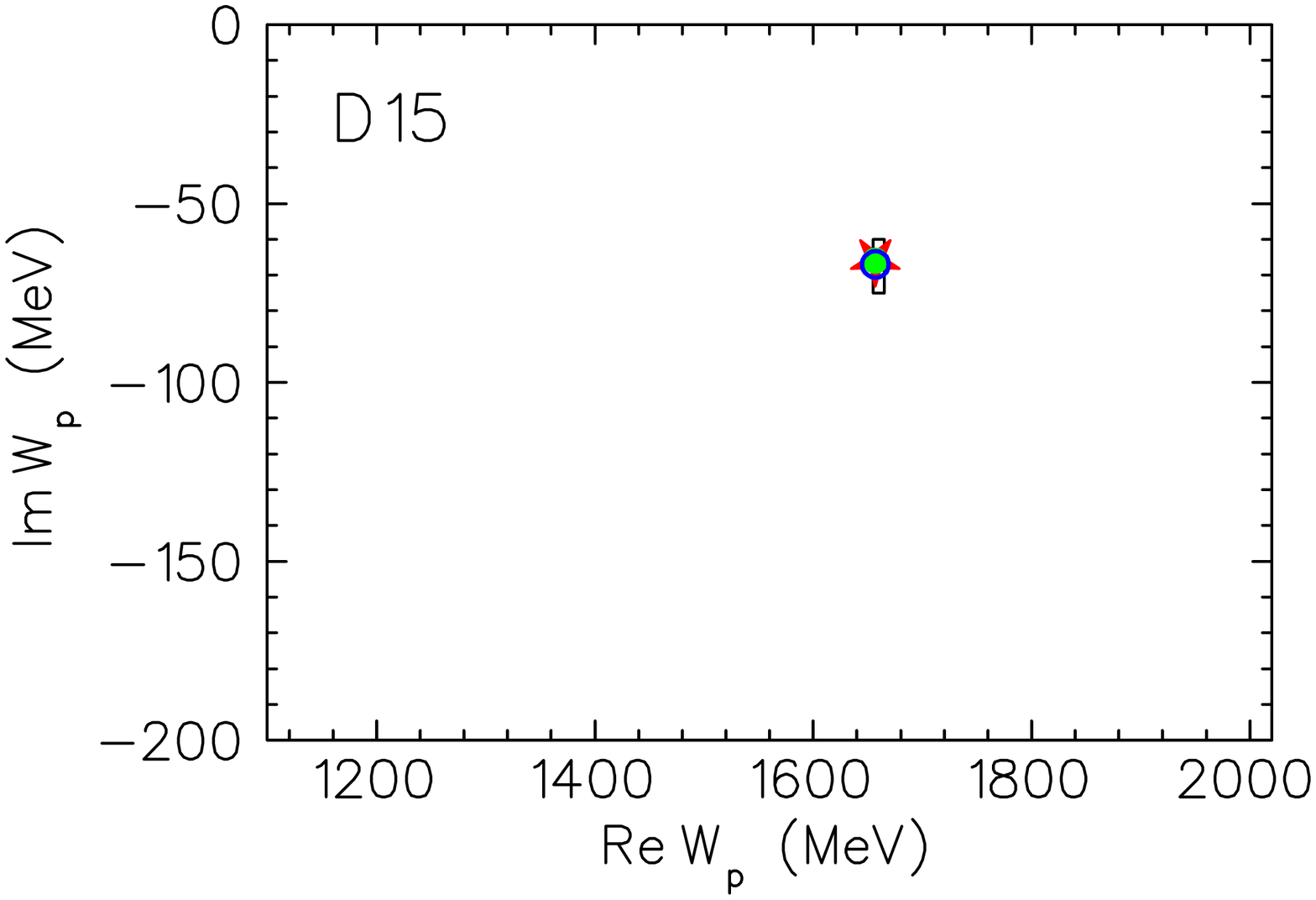}
\includegraphics[width =5.0cm, angle=90]{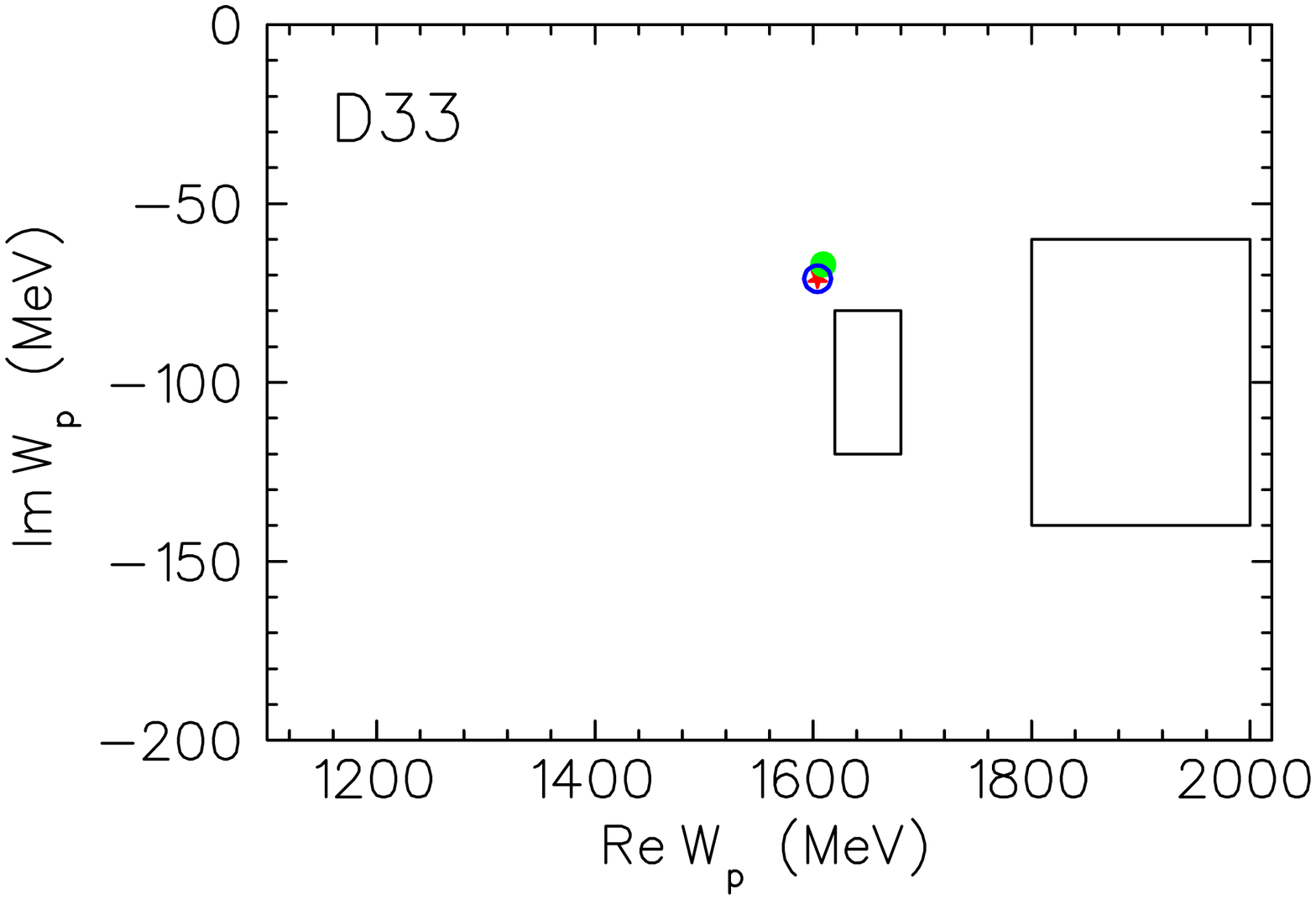}
\hspace*{0.1cm}
\includegraphics[width =5.0cm, angle=90]{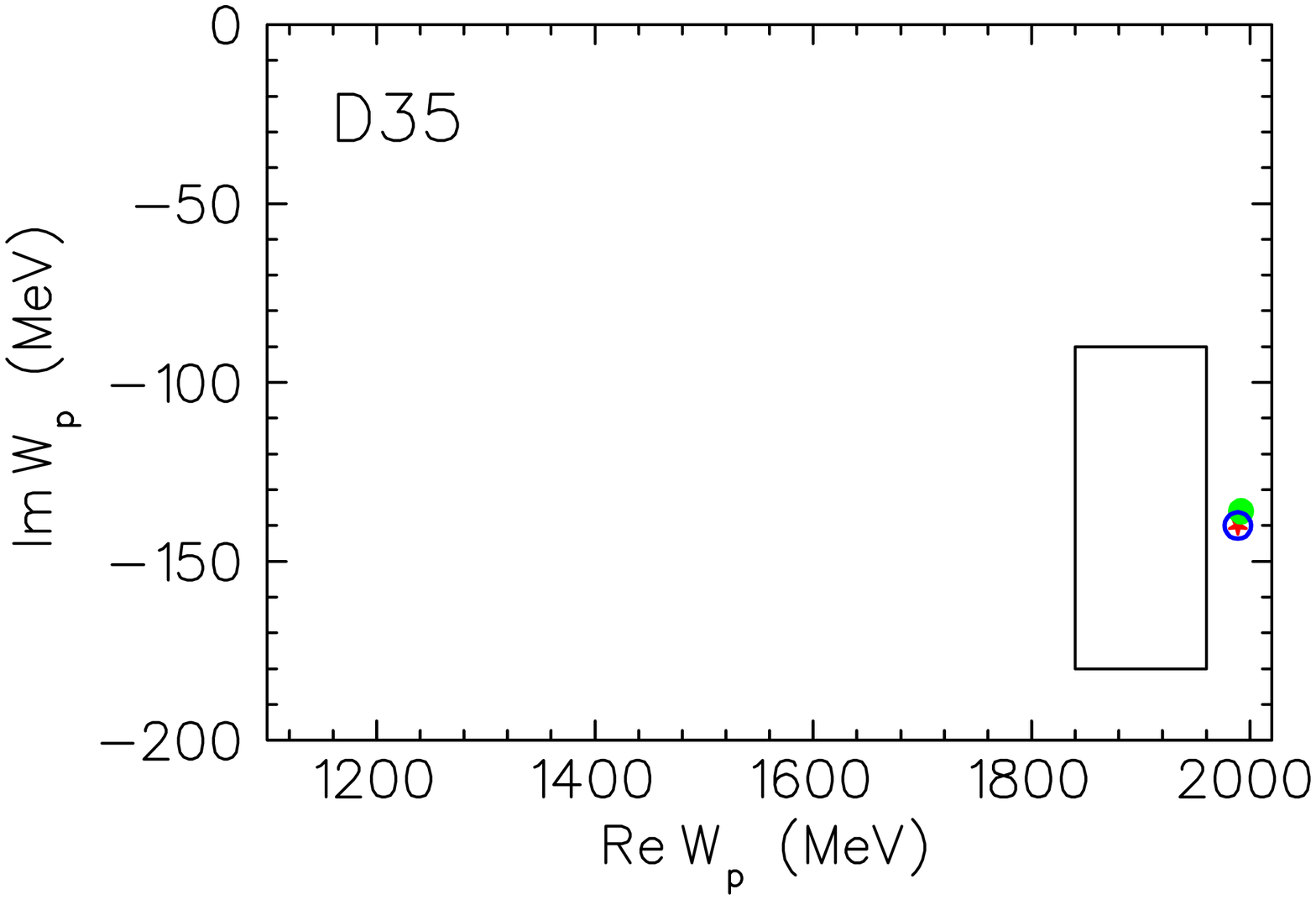}
\end{center}
\caption{The pole positions for the $D$ waves in the complex energy plane. The
notation is the same as in Fig.~\ref{fig:swaves}.} \label{fig:dwaves}
\end{figure}

In Fig.~\ref{fig:dwaves} we show the poles of $D$-wave resonances. PDG lists
three $J=3/2$ states with isospin $1/2$, $D_{13}(1520)^{****}$,
$D_{13}(1700)^{***}$ and $D_{13}(2080)^{**}$, but even a total number of four
regions with extracted pole values, three of these PDG boxes show up in our
figure. Our analysis agrees very well within the reported ranges, where we also
locate four resonance states. While the dominant first resonance is perfectly
described by SP and RM, the SP can only locate another (third) pole. The
improved RM technique, however, is able to find all four positions with high
accuracy.\\

For the $D_{15}$ partial wave we find a rather simple contour, and 2
pole positions can be well located in perfect agreement with
SP and RM. However, only the $D_{15}(1675)^{****}$ lies in our considered range.
\\

For the isospin-3/2 $D$-wave resonances, PDG reports only one 4-star resonance, the
$D_{33}(1700)^{***}$. Our analytic results are reproduced well by SP and
perfectly by RM. A similar agreement is also found for the resonances
$D_{33}(1940)^{*}$ and $D_{35}(11930)^{***}$. All 3 resonances are located
very close to the PDG boxes.
\subsection{F-waves}
\begin{figure}
\begin{center}
\includegraphics[width =5.0cm, angle=90]{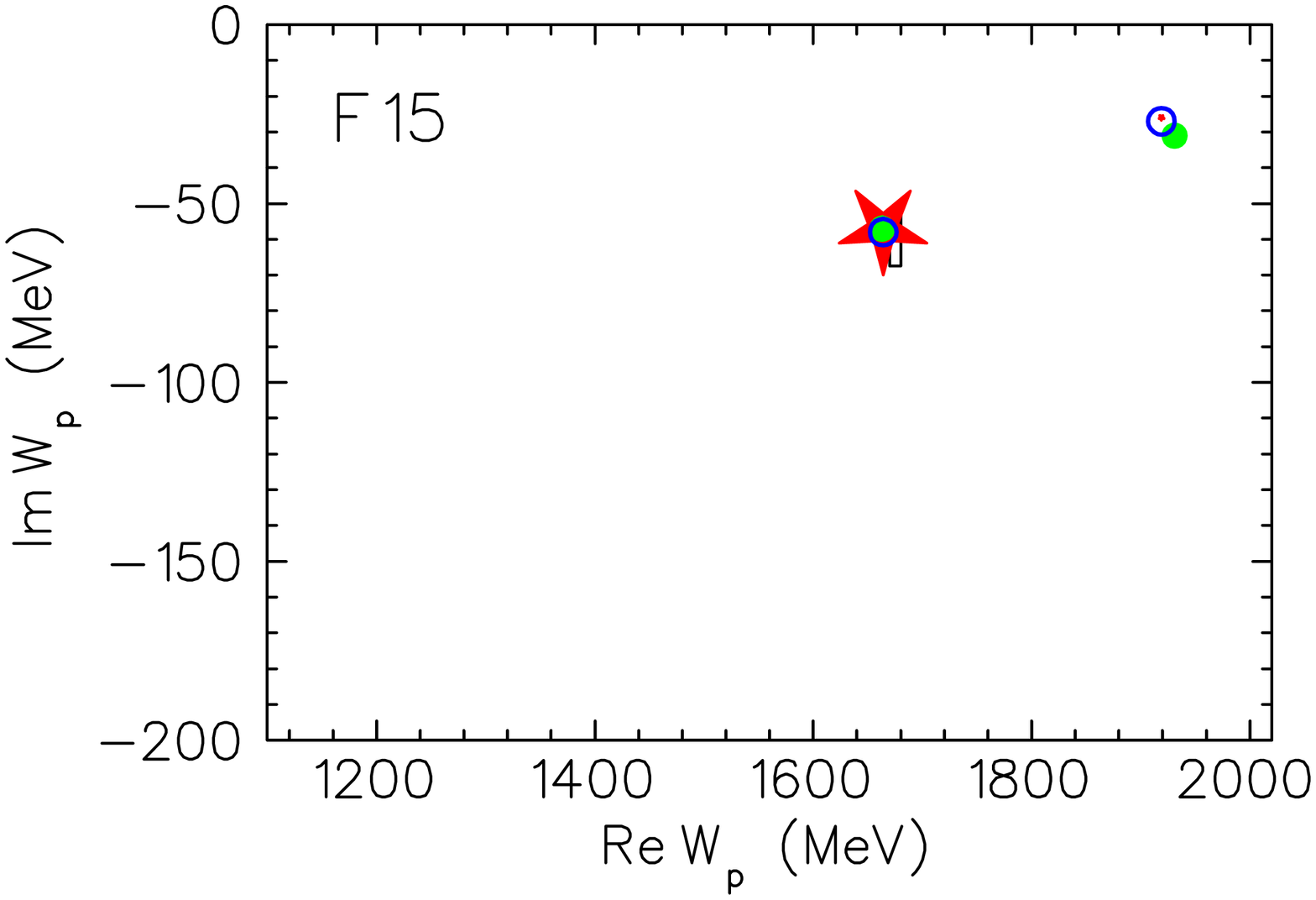}
\hspace*{0.1cm}
\includegraphics[width =5.0cm, angle=90]{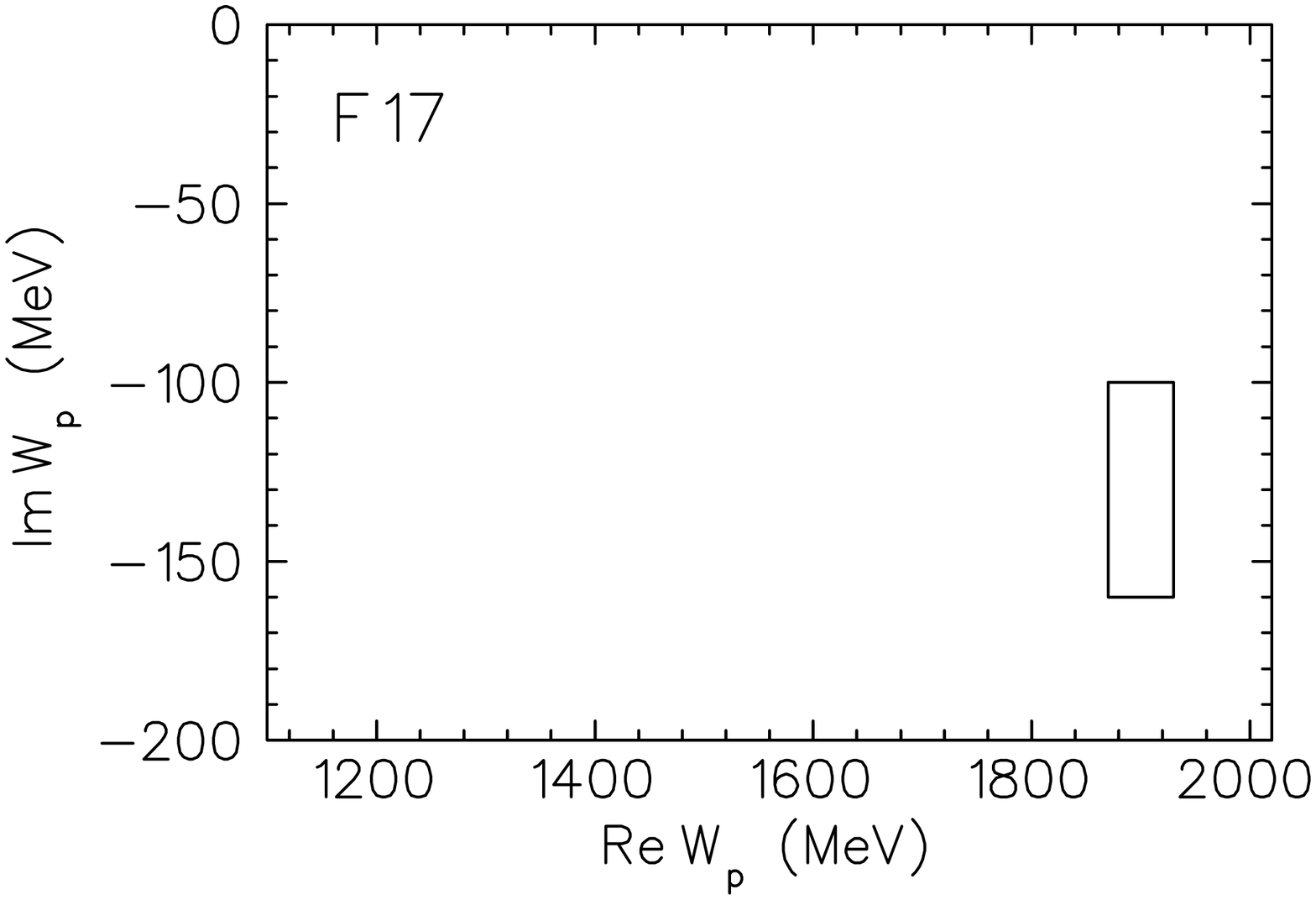}
\includegraphics[width =5.0cm, angle=90]{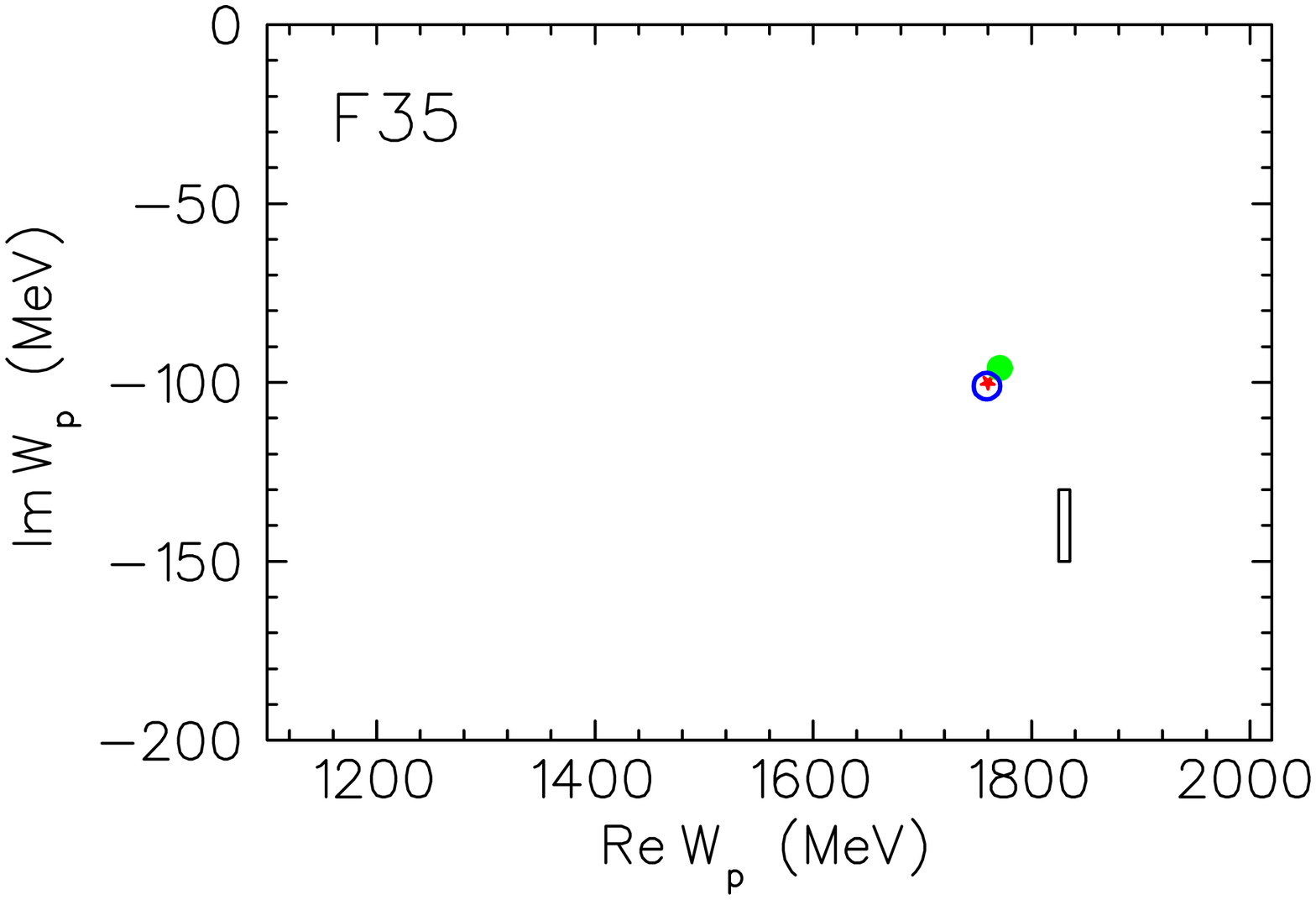}
\hspace*{0.1cm}
\includegraphics[width =5.0cm, angle=90]{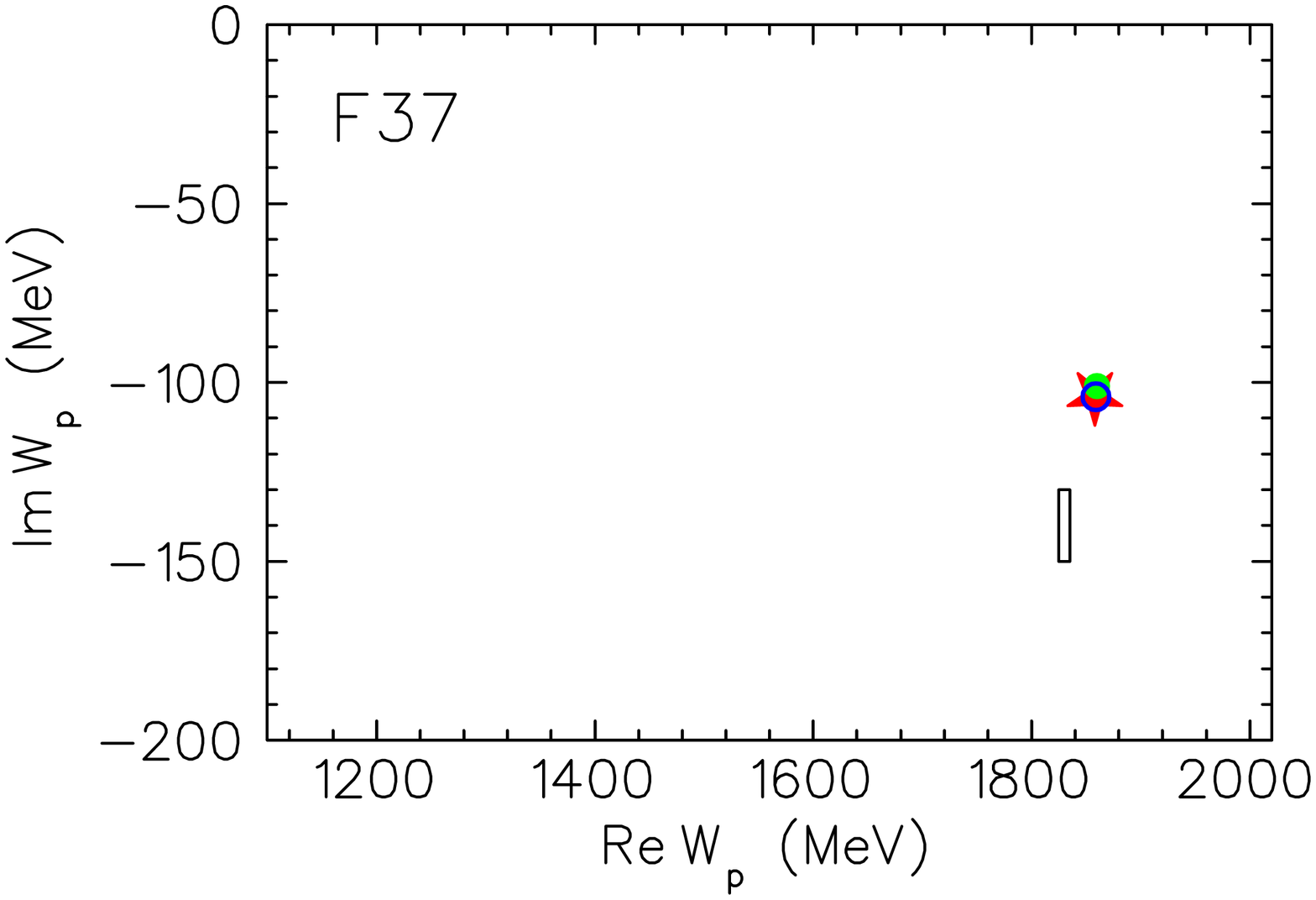}
\end{center}
\caption{The pole positions for the $S$ waves in the complex energy plane.
The notation is the same as in Fig.~\ref{fig:swaves}.} \label{fig:fwaves}
\end{figure}

Figure~\ref{fig:fwaves} displays the $F$-wave resonances. The $F_{15}(1680)^{***}$ is
the most important resonance in the $3^{\rm {rd}}$ resonance region and perfectly
described by SP and RM. We also find a weak second $F_{15}$ state. However,
because of its location close to the real axis, this state is also well described
by SP and RM. In the vicinity, PDG reports the $F_{15}(2000)^{**}$ resonance, however,
with no pole position given. In the $F_{17}$
partial wave we can not find a resonance by our analysis, whereas PDG lists a
$F_{17}(1990)^{**}$. Our result is based on the fact that the inclusion of a
bare resonance in this wave does not significantly improve the $\chi^2$ fit to
the data. A similar conclusion was found in Refs.~\cite{SAID04,Arndt:2006bf}.
\\

For the isospin-3/2 resonances, we find three poles as also reported by PDG.
The first poles in both partial waves,
$F_{35}(1905)^{***}$ and $F_{37}(1950)^{***}$, are well described by SP and RM.
However, our positions are significantly above the PDG bounds situated
at large imaginary values. This disagreement is not too surprising, because
the uncertainty of a pole position rises with the distance from the real axis. We
also find a third pole close to the $F_{35}(2000)^{**}$, which is however very
weak and not well seen in the physical region.
\section{Summary and conclusion}
Within our previously developed Dubna-Mainz-Taipei (DMT) dynamical model we
have studied the pole structure of the pion-nucleon $T$ matrix for all $S$,
$P$, $D$ and $F$ partial waves in the energy range up to c.m. energy
$W=2.0$~GeV. We have analytically continued the $T$ matrix to the unphysical
region in the second Riemann sheet and have accurately calculated the pole
positions and the residues down to ${\rm {Im}} W \geq -200$~MeV. These exact
values were compared with the results obtained by the speed plot, based on the
first derivative of the $T$ matrix, and a recently developed regularization
method using higher derivatives of $T$.
\\

Our general conclusions are:
\begin{itemize}
\item The number and positions of the DMT poles, obtained by the newly developed
analytic continuation method, are in very good agreement with the current PDG results
\cite{PDG2008}.
\item All 4-star resonances are nicely reproduced by both the speed-plot technique and the
renormalization method. In many cases, however, the latter approach gives
significant improvements for the resonance parameters, especially for the pole
residues.
\item We have confirmed the need for a forth S$_{11}$ resonance in addition to the three
resonances listed by PDG~\cite{PDG2008}.  The additional $S_{11}(1878)$ resonance was found
to play an important role in pion photoproduction~\cite{Chen:2002mn},
but was not seen in both the $\pi N \rightarrow\eta N$ reaction and recent data
of $\eta$ photoproduction from the proton~\cite{Thoma:2005awa,Elsner:2007hm}.
\item We have shown that the regularization method is a reliable method to extract
the pole parameters from single-channel data. In the absence of a full experimental
knowledge about all the channels, this method can be sequentially applied in order to track
down the poles relevant for the experimentally known channels of a multichannel amplitude.
\end{itemize}

\vspace{0.5cm}
\centerline{\bf Acknowledgment} S.S.K. wishes to
acknowledge the financial support from the National Science Council
of ROC for his visits to the Physics Department of National Taiwan
University. The work of S.N.Y. is supported in part by the NSC/ROC
under grant No.~NSC095-2112-M022-025. We are also grateful for the
support by the Deutsche Forschungsgemeinschaft through
SFB~443, the joint project NSC/DFG 446 TAI113/10/0-3, and the
joint Russian-German Heisenberg-Landau program.
\\
\newpage
%
% Figures S waves -----------------------------------

%
% Figures P waves -----------------------------------

%
% Figures D waves -----------------------------------

%
% Figures F waves -----------------------------------

%

%----- TABLE 1 ---------------------------------------------------------

\newpage
\begin{table}[htb]
\centering \setlength{\textwidth}{50mm} \caption{Pole positions
$W_p=M_p-\frac{1}{2}i \Gamma_p$ and absolute values of the residues $\mid r_p
\mid$ at the pole, all in MeV, as well as the phases $\theta_p$ of the residues
for $S$-wave resonances. The first lines give the exact pole positions and
residues as obtained by analytic continuation of the DMT model, the second and
third lines show the values obtained by the speed-plot (SP) and the
regularization method (RM(N)) with N being the largest value for stable
derivatives. The star classification and the numerical values of the PDG are
listed in the last line.} \vspace*{0.5cm}
\begin{tabular}{|c|cccc |}
\hline
$N^*$ & $Re W_p$ & $-Im W_p$ & $\mid r_p \mid$ & $\theta_p$(deg)  \\
\hline
$S_{11}(1535)$  & 1499 & 39 & 14 & -45   \\
SP              & 1499 & 26 &  7 & -47   \\
RM(1)           & 1499 & 26 &  7 & -47   \\
**** &$1510\pm 20$& $85\pm 40$ & $ 96\pm 63$ & $ 15\pm 45$  \\
\hline
$S_{11}(1650)$  & 1631 & 60 & 35 & -83   \\
SP              & 1642 & 49 & 22 & -74   \\
RM(6)           & 1631 & 52 & 28 &-119   \\
**** &$1655\pm 15$& $83\pm 8$ & $ 55\pm 15$ & $ -75\pm 25$  \\
\hline
$S_{11}(1880)$  & 1733 & 90 & 16 & -29   \\
SP              &  --  & -- & -- &  --   \\
RM(6)           & 1806 & 83 & 10 &-172   \\
$new$ & & & & \\
\hline \hline
$S_{31}(1620)$  & 1598 & 74 & 23 &  -98  \\
SP              & 1598 & 69 & 23 &  -99  \\
RM(6)           & 1598 & 75 & 24 & -100  \\
**** &$1600\pm 10$& $59\pm 2 $ & $ 16 \pm 3 $ & $-110 \pm 20 $  \\
\hline
$S_{31}(1900)$  & 1774 & 36 & 3.8&  179  \\
SP              & 1775 & 18 & 1.0& -166  \\
RM(5)           & 1777 & 28 & 1  & -157   \\
** &$ 1870 \pm 40$& $90\pm 25$& $10 \pm 3 $ & $ -20 \pm 40 $  \\
\hline\hline
\end{tabular}\label{table:Spoles}
\end{table}
\begin{table}[htb]
\centering \setlength{\textwidth}{50mm} \caption{Pole positions residues for
$P$-wave resonances. For further notation see Tab.~\ref{table:Spoles}.}
\vspace*{0.5cm}
\begin{tabular}{|c|cccc |}
\hline
$N^*$ & $Re W_p$ & $-Im W_p$ & $\mid r_p \mid$ & $\theta_p$(deg)  \\
\hline
$P_{11}(1440)$  & 1371 &  95 & 50 & -79   \\
SP              & 1366 &  90& 48 &  -87  \\
RM(5)           & 1371 &  95& 50 &  -78  \\
**** &$1365\pm 15$& $ 95\pm 15$ & $46\pm 10$ & $-100\pm 35$  \\
\hline
$P_{11}(1710)$  & 1746 & 184& 11 &  -54  \\
SP              & 1721 &  92&  5 & -164  \\
RM(6)           & 1756 & 150& 11 &  -49  \\
*** &$1720 \pm 50 $&$ 115\pm 75$ & $ 10 \pm 4 $ & $-175\pm 35 $  \\
\hline \hline
$P_{13}(1720)$  & 1693 & 136 & 20 & -43 \\
SP              & 1683 & 118 & 15 & -64 \\
RM(4)           & 1695 & 133 & 19 & -34  \\
**** &$1675\pm 15$& $ 98\pm 40$ & $13\pm 7$ & $-139\pm51 $  \\
\hline \hline
$P_{31}(1910)$  & 1900 & 87 & 13 & -116  \\
SP              & 1896 & 68 &  7 & -118  \\
RM(6)           & 1901 & 87 & 10 & -113  \\
**** &$ 1880 \pm 30 $& $100 \pm 20 $ & $20 \pm 4 $ & $-90 \pm 30 $  \\
\hline \hline
$P_{33}(1232)$  & 1212 &  49 & 49 & -42  \\
SP              & 1218 &  44 & 41 & -35  \\
RM(2)           & 1218 &  45 & 41 & -35  \\
**** &$1210\pm 1$& $ 50\pm 1$ & $ 53\pm 2 $ & $ -47 \pm 1 $  \\
\hline
$P_{33}(1600)$  & 1544 &  95&  14& -111  \\
SP              & 1509 &  99&  25& -197  \\
RM(5)           & 1546 &  80&  11& -116  \\
*** &$ 1600\pm 100$& $ 150\pm 50 $ & $ 17\pm 4 $ & $ -150\pm 30$  \\
\hline \hline
\end{tabular}\label{table:Ppoles}
\end{table}
\begin{table}[htb]
\centering \setlength{\textwidth}{50mm} \caption{Pole positions and residues
for $D$-wave resonances. For further notation see Tab.~\ref{table:Spoles}.}
\vspace*{0.5cm}
\begin{tabular}{|c|cccc ||l|cccc |}
\hline
$N^*$ & $Re W_p$ & $-Im W_p$ & $\mid r_p \mid$ & $\theta_p$(deg)  \\
\hline
$D_{13}(1520)$  & 1515 &  60 & 40 &  -7  \\
SP              & 1516 &  61&  40 &  -6  \\
RM(5)           & 1516 &  60&  40 &  -5   \\
**** &$  1510\pm 5  $& $ 55 \pm 5  $ & $ 35 \pm 3 $ & $ -10 \pm 4 $  \\
\hline
$D_{13}(1700)$  & 1718 &  48& 2.8&  -91  \\
SP              &  --  &  --& -- &  --   \\
RM(7)           & 1718 &  49& 2.9&  -91  \\
*** &$ 1680 \pm 50 $& $ 50 \pm 25 $ & $ 6 \pm 3 $ & $ 0 \pm 50 $  \\
\hline
$D_{13}(2080)$  & 1854 & 108& 16 &   -97 \\
SP              & 1834 & 108& 14 &  -134 \\
RM(6)           & 1858 & 105&  9 &   -83 \\
**  &$ 1950 \pm 170 $& $ 100\pm 40 $ & $ 27 \pm 22 $ & $  \sim 0  $  \\
\hline \hline
$D_{15}(1675)$  & 1657 &  66 & 24 & -22  \\
SP              & 1657 &  66&  24 & -23  \\
RM(6)           & 1657 &  67&  25 & -22   \\
**** &$ 1660 \pm 5 $& $ 68 \pm 6 $ & $ 29 \pm 6 $ & $ -30 \pm 10 $  \\
%\hline
%$D_{15}(2200)$  & 2185 & 128& 21 &  -31  \\
%SP              & 2188 & 120& 21 &  -28  \\
%RM(5)           & 2184 & 129& 21 &  -31  \\
% ** &$ 2100 \pm 60 $& $ 180\pm 40 $ & $ 20 \pm 10 $ & $ -90 \pm 50 $  \\
\hline
\hline
$D_{33}(1700)$  & 1604 &  71 & 9.4 & -63 \\
SP              & 1609 &  67&  9.5 & -52  \\
RM              & 1604 &  71&  9.9 & -63   \\
**** &$ 1650 \pm 30 $& $ 100\pm 20 $ & $13 \pm 3 $ & $ -20 \pm 25 $  \\
\hline \hline
$D_{35}(1930)$  & 1989 & 140 & 18 & -78  \\
SP              & 1992 & 136&  19 & -75   \\
RM(5)           & 1989 & 140&  18 & -78   \\
*** &$ 1900 \pm 50 $& $ 133 \pm 48 $ & $18 \pm 6 $ & $ -20\pm 40 $  \\
\hline
\hline
\end{tabular}\label{table:Dpoles}
\end{table}
\begin{table}[htb]
\centering \setlength{\textwidth}{50mm} \caption{Pole positions and residues
for $F$-wave resonances. For further notation see Tab.~\ref{table:Spoles}.}
\vspace*{0.5cm}
\begin{tabular}{|c|cccc ||l|cccc |}
\hline
$N^*$ & $Re W_p$ & $-Im W_p$ & $\mid r_p \mid$ & $\theta_p$(deg)  \\
\hline
$F_{15}(1680)$  & 1664 &  57 & 38 & -26 \\
SP              & 1663 &  57 & 38 & -28 \\
RM(6)           & 1664 &  58 & 39 & -26 \\
**** &$ 1672 \pm 8 $& $ 61 \pm 6 $ & $ 38 \pm 3 $ & $ -23 \pm 7 $  \\
\hline
$F_{15}(2000)$  & 1919 &  26& 1.0  &    15 \\
SP              & 1931 &  31& 1.3  &    89 \\
RM(4)           & 1919 &  27& 1.1  &    15 \\
**              & 1807 &  54.5& 60 & $-67$ \\
\hline \hline
$F_{17}(1990)$  & -- &  --& --&    --  \\
SP              & -- &  --& --&    --  \\
RM              & -- &  --& --&    -- \\
** &$  1900 \pm 30 $& $  260\pm 60  $ & $  9\pm 3    $ & $  -60\pm 30   $  \\
\hline \hline
$F_{35}(1905)$  & 1760 & 100 & 10 & -66  \\
SP              & 1771 &  96 & 11 & -47  \\
RM(5)           & 1759 & 101 & 10 & -66   \\
**** &$ 1830 \pm 5 $& $ 140 \pm 10 $ & $ 25\pm 8 $ & $ -50\pm 20 $  \\
\hline \hline
$F_{37}(1950)$  & 1858 & 104 & 43 & -48  \\
SP              & 1860 & 101&  44 & -45   \\
RM(5)           & 1859 & 104&  44 & -47   \\
**** &$ 1880 \pm 10 $& $ 140 \pm 10 $ & $ 50\pm 7 $ & $ -33 \pm 8 $  \\
\hline
\hline
\end{tabular}\label{table:Fpoles}
\end{table}
\end{document}